\documentclass[%
 reprint,
superscriptaddress,
 amsmath,amssymb,
prb,
]{revtex4-1}

\usepackage{graphicx}
\usepackage{dcolumn}
\usepackage{bm}
\usepackage{comment}

\usepackage{braket}
\usepackage{color}
\usepackage{array}
\newcolumntype{?}{!{\vrule width 1pt}}

\begin{document}

\preprint{APS/123-QED}

\title{{Single-Particle-Picture Breakdown in laterally weakly confining GaAs Quantum Dots}}

\author{Daniel Huber}
\email{daniel.huber@jku.at}
\affiliation{Institute of Semiconductor and Solid State Physics, Johannes Kepler University, Linz, Altenbergerstr. 69, 4040, Austria}
\affiliation{Secure and Correct Systems Lab, Linz Institute of Technology, Linz, Altenbergerstr. 69, 4040, Austria}

\author{Barbara Ursula Lehner}%
	\affiliation{Institute of Semiconductor and Solid State Physics, Johannes Kepler University, Linz, Altenbergerstr. 69, 4040, Austria}
	\affiliation{Secure and Correct Systems Lab, Linz Institute of Technology, Linz, Altenbergerstr. 69, 4040, Austria}
 
\author{Diana Csontosov\'{a}}%
    \affiliation{Department of Condensed Matter Physics, Faculty of Science, Masaryk University, Kotl\'a\v{r}sk\'a~267/2, 61137~Brno, Czech~Republic}
    \affiliation{Czech Metrology Institute, Okru\v{z}n\'i 31, 63800~Brno, Czech~Republic}
  
\author{Marcus Reindl}%
	\affiliation{Institute of Semiconductor and Solid State Physics, Johannes Kepler University, Linz, Altenbergerstr. 69, 4040, Austria}

\author{Simon Schuler}%
    \affiliation{Institute of Semiconductor and Solid State Physics, Johannes Kepler University, Linz, Altenbergerstr. 69, 4040, Austria} 
    
\author{Saimon Filipe Covre da Silva}%
    \affiliation{Institute of Semiconductor and Solid State Physics, Johannes Kepler University, Linz, Altenbergerstr. 69, 4040, Austria}
    
\author{Petr Klenovsk\'{y}}%
    \email[]{klenovsky@physics.muni.cz}
    \affiliation{Department of Condensed Matter Physics, Faculty of Science, Masaryk University, Kotl\'a\v{r}sk\'a~267/2, 61137~Brno, Czech~Republic}
    \affiliation{Czech Metrology Institute, Okru\v{z}n\'i 31, 63800~Brno, Czech~Republic}
       

    
    
    
    
\author{Armando Rastelli}%
\email{armando.rastelli@jku.at}
	\affiliation{Institute of Semiconductor and Solid State Physics, Johannes Kepler University, Linz, Altenbergerstr. 69, 4040, Austria}






\begin{abstract}
We present a detailed investigation of different excitonic states weakly confined in single GaAs/AlGaAs quantum dots obtained by the Al droplet-etching method. For our analysis we make use of temperature-, polarization- and magnetic field-dependent $\mu$-photoluminescence measurements, which allow us to identify different excited states of the quantum dot system. Besides that, we present a comprehensive analysis of g-factors and diamagnetic coefficients of charged and neutral excitonic states in Voigt and Faraday configuration. Supported by theoretical calculations by the Configuration interaction method, we show that the widely used single-particle Zeeman Hamiltonian cannot be used to extract reliable values of the g-factors of the constituent particles from excitonic transition measurements.
 \end{abstract}

\pacs{Valid PACS appear here}

\maketitle

\section{Introduction}

Within the last years GaAs quantum dots (QDs), obtained with the droplet-etching method\cite{Wang2007,Heyn2009,Huo:APL2013}, emerged as a promising source of non-classical states of light, such as single photons with a strongly suppressed multi-photon emission probability \cite{doi:10.1063/1.5020038}, highly indistinguishable photon states \cite{Huber2017,Reindl2019,doi:10.1021/acs.nanolett.8b05132,Liu2019}, and single polarization entangled photon-pairs with a near unity degree of entanglement \cite{Huber2017,Keil2017,Huber2018,Gurioli2019}. 

Only recently it was realized that excitons must be weakly confined in these QDs, as the measured groundstate exciton (X) lifetimes of about 250 ps \cite{Huber2017,Huber2018,Keil2017} are significantly lower than the minimum lifetime expected for a strongly confining system (440 ps) \cite{PhysRevB.82.233302,Reindl2019}. The morphology of such GaAs/AlGaAs quantum dots (QDs) features in fact a lateral extension to be much larger than the free exciton Bohr radius in GaAs \cite{Huo:APL2013,Huber2017,Reindl2019}. Therefore, the excitonic states are laterally weakly confined and the Coulomb interaction between the charge carriers is supposed to overwhelm quantum confinement effects. In turn, the weak confinement regime is partly responsible for the excellent optical properties of GaAs QDs, as the short lifetime and large in-plane extension of the wavefunction limit the influence of dephasing and structural anisotropies within the QD~\cite{Huber:rev2018}.

We note that, a standard model system for the weak confinement regime is represented by GaAs QDs formed at thickness fluctuations in thin GaAs/AlGaAs quantum wells \cite{PhysRevLett.95.067401}, which, however, suffer from a poor control of the lateral confinement. Furthermore, the energy separation between confined states and delocalized states is small. In contrast to that, droplet-etched GaAs QDs avoid these issues and provide an ideal system to study the properties of weakly confined excitons. 

Besides possible applications of QDs as single photon and entangled photon-pair sources, the spin states of charges confined in a QD may serve as qubits for quantum technology~\cite{PhysRevA.57.120,PhysRevX.5.011009}. Moreover, semiconductor QDs could provide a link between photonic and spin qubits via photon-to-spin conversion \cite{Gaudreau_2017}. Such complex applications require a detailed knowledge about the response of an excited state in the QD to an external magnetic field, which is described by its g-factor and the diamagnetic coefficient. Furthermore, an individual engineering of the contribution of electrons ($e^-$) and holes ($h^+$) within a complex to the overall g-factor is desired \cite{PhysRevB.99.195305,Bennett2013,Medeiros-Ribeiro2003}. Hence, for engineering the magnetic properties of a QD precise measurements are crucial. Usually, to extract the g factors of $e^-$ and $h^+$ confined in single QDs from photoluminescence (PL) measurements a single-particle (SP) Zeeman Hamiltonian~\cite{PhysRevB.65.195315} is employed~\cite{Bennett2013,PhysRevB.94.245301,PhysRevB.99.195305,doi:10.1063/1.3367707} (for details about the method we refer the interested reader to Ref. \cite{PhysRevB.84.195305}). However, it is questionable if one can apply such an approach to weakly confining QDs. The knowledge of the magneto-optical properties of GaAs/AlGaAs QDs obtained by droplet etching is restricted to only a few works~\cite{Huo:NatPhys,PhysRevB.93.165306,lobl2019} and a comprehensive analysis is to the best of our knowledge missing.

In this work we first study various excitonic transitions in GaAs QDs via polarization-resolved and temperature-dependent $\mu-$PL measurements, which allow us to identify several charged states such as the positive (X$^+$) and negative (X$^-$) trion as well as some of their excited states. We then present a comprehensive study on the magneto-optical properties of the GaAs/AlGaAs QD system. We apply magnetic fields along the QD growth direction [001] (Faraday configuration) and along the [110] direction (Voigt configuration) and extract the diamagnetic coefficients and g-factors of several excited states in the GaAs QDs. To gain more insight into the physical properties of our QDs under an external magnetic field, we complement our experimental study with calculations using the method of Configuration interaction (CI)~\cite{Takagahara1993,SFZ01,Schliwa:09,Klenovsky2017}. The theoretical results are in good agreement with the experimental data and confirm that correlation effects among the confined carriers have significant influence on the magnetic properties. Finally, we demonstrate by experiment and CI calculations that a SP picture, which turns out to be adequate in the case of strongly confining QDs, leads to poor results in a weakly confining system \cite{PhysRevB.65.195315,PhysRevB.84.195305,Kunz2013}. 

\section{Experimental Results}

\subsection{Polarization-resolved and temperature-dependent $\mu-$PL measurements}

\begin{figure}[htbp]
	\centering
		\includegraphics[width=85mm]{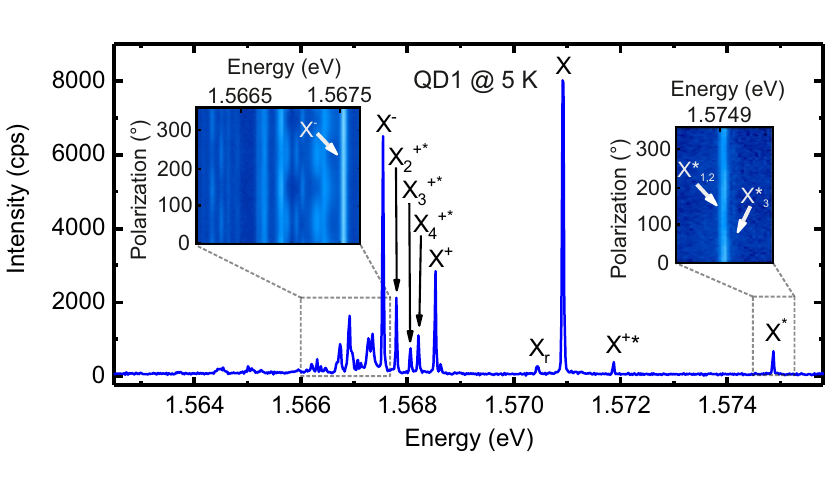}
	\caption{Spectrum of a representative quantum dot (QD1) under above bandgap excitation at 5 K. The identified recombination channels are labeled within the figure, see text for details. The insets show color-coded $\mu$-PL spectra of transitions within the grey dotted boxes.} 
	\label{fig:fig2}
\end{figure}

We begin by characterizing the optical transitions of our GaAs/AlGaAs QDs. The details on the sample growth are given in Ref. \cite{Huber2018}. For the excitation we use a 532 nm continuous wave laser, which is focused on the sample through an aspheric lens with 0.65 numerical aperture. The low QD density ($\sim2\times10^7$~cm$^{-2}$) allows us to address the emission of single QDs. In above bandgap excitation, $e^-$ and $h^+$ are mainly generated in the barrier material of the QD, where the carriers subsequently diffuse, partly get trapped by the QD confinement potential, and relax to the low energy levels. The spectrum of a representative QD is shown in Fig.~\ref{fig:fig2}.
The spectral position of X within the cluster of lines is well known from previous experiments \cite{Huo:APL2013,Huber2017,Huber2018,Reindl2019}. It has two orthogonally polarized components, which are typically non-degenerat in energy due to the fine structure splitting \cite{PhysRevB.65.195315}. The biexciton state is not visible under the used excitation conditions, which we attribute to competition with other configurations due to a slow relaxation of carriers to the ground state via multiple acoustic phonons \cite{Huber2017,Reindl2019}.

Above-bandgap excitation provides an interesting feature to identify some of the observed transitions: The mobility of $e^-$ and $h^+$ within the AlGaAs barrier influences the formation probability of the different charge complexes. The mobility is reduced due to scattering events mainly with ionized impurities and phonons, such that the larger scattering cross section of the $h^+$ compared to the $e^-$ plays a significant role in the spectral response \cite{PhysRevB.73.115322}. We demonstrate this claim by performing a temperature-dependent $\mu-$PL measurement. By increasing the temperature the mobility of the $e^-$ increases compared to the one for the $h^+$ and we expect that $h^+$-dominated complexes (e.~g. X$^+$) are less often formed than $e^-$-dominated ones (e.~g. X$^-$).

\begin{figure}[htbp]
	\centering
		\includegraphics[width=75mm]{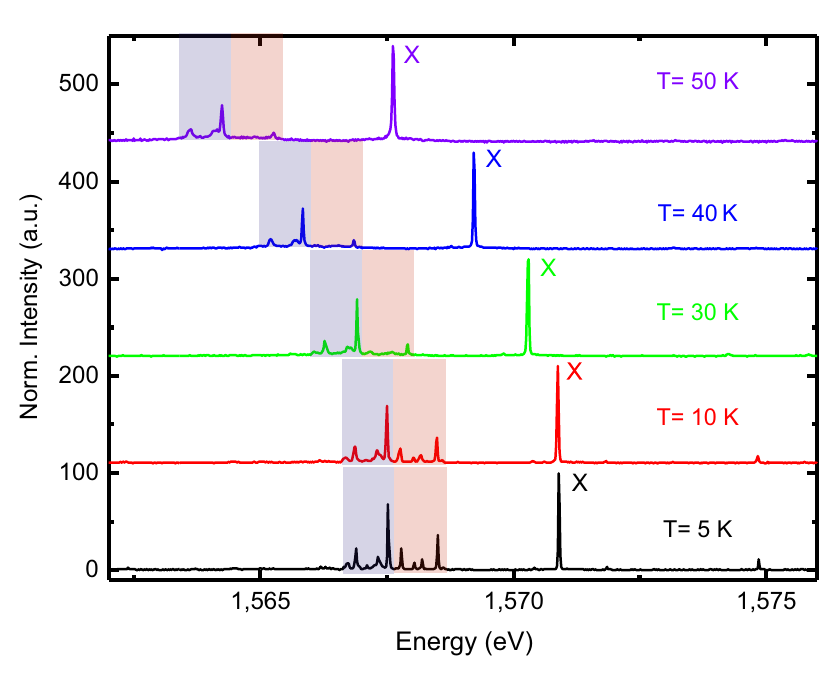}
	\caption{Normalized temperature-depended $\mu$-PL spectra of a representative quantum dot (QD1). The quantum dot is excited by a 532 nm continuous wave diode laser above the AlGaAs barrier bandgap.} 
	\label{fig:fig1}
\end{figure}

Fig.~\ref{fig:fig1} shows temperature-dependent $\mu$-PL spectra for temperatures between 5 K and 50 K.
The spectra of a representative QD (QD1) are normalized with respect to the intensity of the X transition. On the low energy side of the X, transitions with energies $>$ 1.5675 eV (see red area in Fig.~\ref{fig:fig1}) drop fast in intensity with increasing temperature until they completely disappear. In contrast, transitions with energy $<$ 1.5675 eV (see blue area in Fig.~\ref{fig:fig1}) are less sensitive to a temperature change.

By combining the temperature-dependence with polarization resolved $\mu-$PL measurements we can now discuss the origin of several transitions in Fig. \ref{fig:fig2}. While the SP picture is poorly suitable for describing the excitonic states of weakly confining QDs \cite{Reindl2019}, we stick to that for the moment to provide an intuitive description of the charge configuration of the individual states. On the low energy side of X ($<1.570918$ eV for QD1 at 5 K) we first find a transition labeled as X$_{r}$. 
This feature consists of two linearly-polarized lines with an energy splitting and polarization orientation equal to those of X and disappears under quasi-resonant excitation. We thus attribute it to a quantized energy jitter of the X stemming from charging and uncharging of a defect in the vicinity of the QD. At an energy of 1.568524 eV we then find the X$^+$ transition and at 1.567542 the X$^-$. We can distinguish between those by the different temperature trend seen in Fig.~\ref{fig:fig1}. The trions have no measurable polarization splitting as expected for a Kramers doublet~\cite{doi:10.1002/pssa.200306157}. Between the trions we observe three additional transitions, which are, according to the temperature-dependent measurements, $h^+$ dominated. We thus attribute these lines to the excited states of $X^+$ (hot trions) \cite{PhysRevB.81.245304,PhysRevB.79.125316,doi:10.1002/pssa.200306157,PhysRevB.73.115322}. In the SP picture, the simplest hot trion is described by an $e^--h^+$ pair in the s-shell and an extra $h^+$ ($e^-$) in the p-shell. We want to point out that in a more realistic picture including carrier interactions the situation is not that trivial (see Supplementary VI). In most of the QD systems it is difficult to observe radiative transitions from an excited trion as that is converted rapidly to a ground state trion via non-radiative relaxation. However, the slow relaxation rates between the energy levels in our QDs allow us to study these transitions in more detail~\cite{Reindl2019}. Due to exchange interaction, the excited positive trion splits into four degenerated doublets \cite{doi:10.1002/pssa.200306157,PhysRevB.81.245304}:
\begin{equation}
        \ket{\rm X_{4}^{+*}}= \begin{cases} \uparrow_{s}(\Uparrow_{s}\Downarrow_{p}-\Downarrow_{s}\Uparrow_{p}) & J_z=+1/2 \\ \downarrow_{s}(\Uparrow_{s}\Downarrow_{p}-\Downarrow_{s}\Uparrow_{p}) & J_z=-1/2 \end{cases}
\label{eq:1}
\end{equation} 
\begin{equation}
      \ket{\rm X_{3}^{+*}}= \begin{cases} \uparrow_{s}(\Uparrow_{s}\Downarrow_{p}+\Downarrow_{s}\Uparrow_{p}) & J_z=+1/2 \\  \downarrow_{s}(\Uparrow_{s}\Downarrow_{p}+\Downarrow_{s}\Uparrow_{p}) & J_z=-1/2  \end{cases}
      \label{eq:2}
\end{equation} 
\begin{equation}
      \ket{\rm X_{2}^{+*}}=  \begin{cases}  \downarrow_{s}\Uparrow_{s}\Uparrow_{p} & J_z=(+5/2) \\ \uparrow_{s}\Downarrow_{s}\Downarrow_{p} & J_z=(-5/2) \end{cases}
\label{eq:3}
\end{equation} 
\begin{equation}
\ket{\rm X_{1}^{+*}}= \begin{cases} \uparrow_{s}\Uparrow_{s}\Uparrow_{p} & J_z=(+7/2)\\ \downarrow_{s}\Downarrow_{s}\Downarrow_{p} & J_z=(-7/2) \end{cases}
\label{eq:4}
\end{equation}
where $\uparrow_{i},\downarrow_{i}$ ($\Uparrow_{i},\Downarrow_{i}$) describe the $e^-$ ($h^+$) spin configuration in the shell $i\in\{\rm s,\rm p\}$ and the number in the brackets gives the total angular momentum projection on the quantization axis under the simplified assumption that holes in the s- and p-states have pure heavy-hole (HH) character (with $J_z=\pm 3/2$). The singlet state (X$_{4}^{+*}$) and two of the triplet states (X$_{3}^{+*}$ and X$_{2}^{+*}$) emit a single photon when the s-shell $e^--h^+$ pair recombines, while the remaining triplet state (X$_{1}^{+*}$) is forbidden due to dipole selection rules. Since the energetic ordering of X$_{2}^{+*}-$X$_{4}^{+*}$ is non trivial, we follow the labeling given in Ref. \cite{PhysRevB.81.245304,PhysRevB.79.125316}.


Below the X$^-$ we observe several transitions, which, according to the temperature trend, are $e^-$ dominated. Nevertheless, a detailed analysis of these states is not possible in our measurements due to the limited spectral resolution ($\approx 25$ $\mu$eV). A polarization resolved $\mu-$PL measurement (see inset Fig. \ref{fig:fig2}) shows non-polarized lines as well as doublets of orthogonally polarized lines. We speculate that these states belong to the excited X$^-$ and/or to multiple negatively charged states.

At energies above the X transition we observe two transitions (X$^{+*}$ and X$^*$), which disappear with increasing temperature (see Fig. \ref{fig:fig2}). The X$^{+*}$ at 1.571869 eV does not show any polarization splitting. We conclude that this is a transition of an excited $X^+$ with, where the $h^+$ in the p-shell recombines with an $e^-$ in the s-shell. The  X$^*$ is a complex of three transitions (see inset Fig. \ref{fig:fig2}), where two (X$^*_{1,2}$ 1.574865 eV) have similar intensities and are orthogonally polarized with an energy splitting of $5\,\mu$eV. The third transition (X$^*_3$ at 1.574931 eV) is linearly polarized and by a factor $\sim15$ lower in intensity. This is an excited complex of the bright X doublet (total angular momentum of $\pm 1$) and one component of the dark X doublet (total angular momentum of $\pm 2$), whereby the $h^+$ of the complex is situated in the p-shell. According to our calculations (discussed later on), we attribute the brightening of one of the nominally dark states to the fact that the p-state holes have $20 \%$ light-hole (LH) character, which allows for a weak dipole transition (see Supplementary Section I)~\cite{Huo:NatPhys}. For clarity, we label the whole transition complex with $X^*$ instead of labeling each transition individually.      

\subsection{Magneto-optical properties of GaAs QDs}

We continue our study by investigating the magnetic response of the excitonic states discussed above. The QD sample is mounted in a He bath cryostat equipped with a superconducting vector magnet. In Faraday configuration, the magnetic field vector is aligned along the [001] crystal axis (growth direction), which we label as $z$. Due to the in-plane symmetry of the QDs, we restrict our measurements in Voigt configuration to magnetic fields along [110] crystal axis only, which we label as $x$. The magnetic field alters the emission properties of the QD, where (i) the diamagnetic shift and (ii) the Zeeman effect are the dominant processes.

In (i) the magnetic field induces a magnetic moment and changes the energy of a state in first approximation according to: 
\begin{equation}
    \Delta E=  \gamma B^2,
\label{eq:dia}
\end{equation}
where $\gamma$ is the diamagnetic coefficient. For the neutral exciton, the diamagnetic coefficient probes the spatial extent of the excitonic wavefunction, which depends on the spatial confinement and interactions between the confined particles \cite{PhysRevB.57.6584,PhysRevB.66.193303,PhysRevLett.101.267402}. Hence, it is obvious that the diamagnetic shift is a direction-dependent quantity. The parabolic behaviour from Eq.~(\ref{eq:dia}) is only valid in the weak-field limit, where the magnetic length $l_{M}=\sqrt{\frac{\hbar}{eB}}$ is larger than the spatial extent of the wavefunction $l_{wf}$ \cite{PhysRevB.81.113307}. Note that the magnetic length is $l_{M}\approx 15 $ nm at a magnetic field of 3 T, while the excitonic wavefunction may exceed this value in  our GaAs QDs (the QDs have a base diameter of $\approx 60 $ nm and a height of $\approx 10$ nm), such that the diamagnetic shift can deviate from the $B^2$ dependence.

In (ii) a magnetic field along $z$ lifts the spin degeneracy, while a field along $x$ also breaks the symmetry of the system introducing a coupling between different states (for example between dark and bright X) \cite{PhysRevB.65.195315,PhysRevB.84.195305}. The relation between magnetic field and splitting (mixing) is characterized by the g-tensor, whereby we probe its elements along $x$ ($g_{x}$) and $z$ ($g_{z}$)  \cite{PhysRevB.84.195305}. 

The Zeeman effect in QDs is commonly described by a SP Zeeman Hamiltonian, where the g-factor is a combination of $e^-$ and $h^+$ g-factors~\cite{PhysRevB.65.195315,PhysRevB.84.195305}. However, we find in the following that in the weak confinement regime this approximation is not valid.

\begin{figure}[htbp]
	\centering
		\includegraphics[width=90mm]{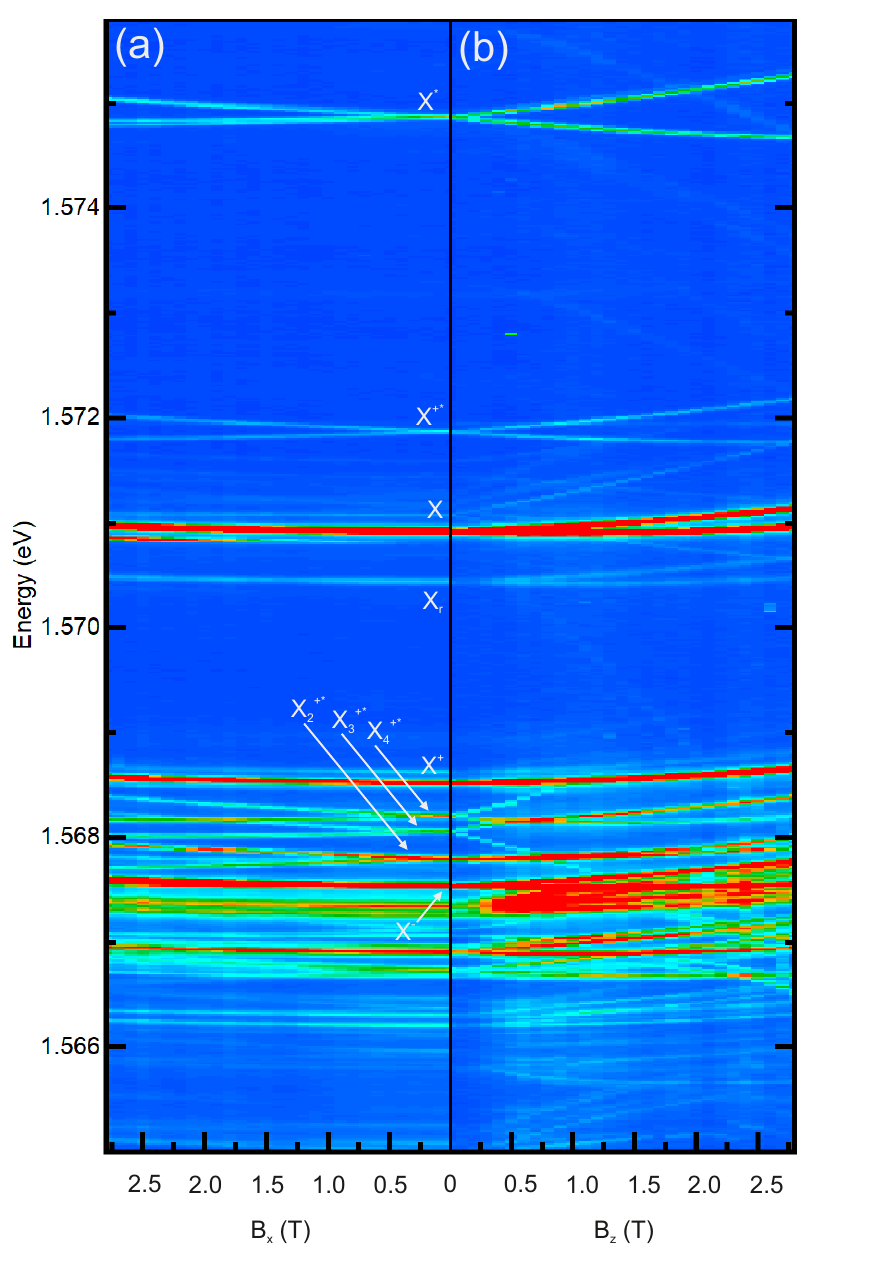}
	\caption{$\mu$-PL spectra under a magnetic field of a representative QD (QD1) applied along (a) [110] and (b) [001] crystal direction.} 
	\label{fig:fig3}
\end{figure}

Fig.~\ref{fig:fig3} shows the shift of the transition lines of QD1 versus the magnetic field between 0 and 2.5~T applied in $x$- and $z$-direction, respectively. 
To get full insight into the magnetic properties of the different transitions we additionally record polarization-resolved spectra versus magnetic field strength ($B$). Finally, we use these data to extract the diamagnetic coefficients and g-factors by fitting the energy shifts with the following model \cite{PhysRevB.85.165323}:  
\begin{equation}
    E_{\uparrow/\downarrow}=E_{0}+\gamma B^2\pm \frac{1}{2} \sqrt{S^2+(g_{0}+g_{2}B^2)^2\mu_{B}^2 B^2},
\label{eq:gfactor}
\end{equation} 
where $\uparrow/\downarrow$ label the two Zeeman-split states, $E_0$ is the energy of the transition at $B=0$, $g_{0}$ is the g-factor, $g_{2}$ is a second order term, $S$ a possible initial fine-structure splitting and $\mu_{B}$ is the Bohr magneton. For convenience, we separate the diamagnetic shift and the Zeeman interaction during fitting by using $\frac{1}{2}(E_{\uparrow}+E_{\downarrow})$ to obtain $\gamma$ and $E_{\uparrow}-E_{\downarrow}$ to obtain $g_0$ and $g_2$, respectively. In Faraday configuration, the model in Eq.~(\ref{eq:gfactor}) can be used to fit satisfactorily all transitions. However, in Voigt geometry the situation is more complicated due to mixing between the states.
We can still use Eq.~(\ref{eq:gfactor}) to fit the X transition for small fields, where the bright-dark mixing is negligible, and also for X$_{2}^{+*}-$X$_{4}^{+*}$, each splitting into two circularly polarized components. The X$^+$ and X$^-$, which are not much influenced by the exchange interaction, split instead into four linearly polarized components each, and thus cannot be described by Eq.~(\ref{eq:gfactor}). Therefore we follow the approach outlined in Ref.~\cite{Bennett2013} and calculate the g-factor according to: 
\begin{equation}
    g_x=\frac{E_1-E_2}{\mu_B B},
\label{eq:gfactor2}
\end{equation} 
where $E_1$ ($E_2$) is the highest (lowest) energy component of the trion quadruplet. We use the same model to extract the g-factors of X$^{+*}$ and X$^{*}$, which are found to split into four linearly polarized components. In Tab.~\ref{tab:tab1} we summarize the results for a representative QD (QD1). 
\begingroup
\squeezetable
\begin{table*}[htbp]
\caption{The table summarizes fitted diamagnetic coefficients and g-factors for the excitonic transitions in a representative quantum dot (QD1) and the CI calculation results for a magnetic field applied along [110] ($x$) direction and [001] ($z$) direction.}
\begin{ruledtabular}
 \begin{tabular}{|c|c|c|c|c|c|c||c|c|c|c|c|c|c|}
 \hline  QD1  & $\gamma_{x}$ ($\frac{\mu eV}{T^2}$) & $g_{x}$ & $g_{2,x}$ ($T^{-2}$) & $\gamma_{z}$ ($\frac{\mu eV}{T^2}$) & $g_{z}$ & $g_{2,z} (T^{-2})$  & CI & $\gamma_{x}$ ($\frac{\mu eV}{T^2}$) & $g_{x}$ & $g_{2,x}$ ($T^{-2}$) & $\gamma_{z}$ ($\frac{\mu eV}{T^2}$) & $g_{z}$ & $g_{2,z} (T^{-2})$ \\  \hline 
 
  X$^*$ & 5.9 (5) & 1.61 (3) & - & 12.7 (3) & 3.74 (1) & -0.01 (1)  
  & X$^*$ & 4.56 & 1.7 & 0.2 & 14.7 & 2.57 & 0.74 \\ \hline
  X$^{+*}$ & 6.4 (5) & 1.45 (3) & - & 11.5 (9) & 2.58 (1) & -0.01 (1)  & X$^{+*}$ & 6.92 & 0.46 & 0.03  & 8.03 & 0.58 & 0.12 \\ \hline
  X & 6.4 (1) & 0.03 (1) & 0.002 (1) & 16.81 (7) & 1.112 (1) & -0.001 (1) &
  X & 6.46 & 0.45 & 0.16 & 15.6 & 1.25 & 0.03 \\ \hline
  X$^+$ & 7.6 (7) & 0.18 (5) & - & 16.29 (4) & 0.13 (1) & 0.001 (1) &
  X$^+$ & 5.8 & 0.48 & 0.01& 16.17 & 0.59 & 0.12 \\ \hline
  X$_{4}^{+*}$ & 10.4 (3) & 1.41 (1) & 0.003 (1) & 139 (1) & 6.33 (1) & 0.18 (1) &
  X$_{4}^{+*}$ & 4.17 & 1.69 & 0.04 & 96.11 & 2.93 & 0.87 \\ \hline
  X$_{3}^{+*}$ & 10.6 (9) & 1.13 (5) & 0.18 (4) & -94 (1) & 8.27 (1) & -0.33 (1) &
  X$_{3}^{+*}$ & 2.07 & 1.7 & 0.07 & -55.32 & 3.31 & 1.03 \\ \hline
  X$_{2}^{+*}$ & 6.8 (3) & 1.3 (1) & 0.01 (1) & 20.6 (1) & 0.27 (1) & 0.04 (1) &
  X$_{2}^{+*}$ & 7.29 & 1.66 & 0.06 & 10.59 & 0.64 & 0.08 \\ \hline
  X$^-$ & 6 (1) & 0.33 (3) & - & 16.9 (5) & 1.42 (1) & -0.046 (1) &
   X$^-$ & 4.46 & 0.919 & 0.0009 & 15.61 & 2.12 & 0.06 \\ \hline
 \end{tabular}
 \end{ruledtabular}
\label{tab:tab1}
\end{table*}
\endgroup

We observe pronounced anisotropies of the g-factors along $x$ and $z$-direction. This phenomenon has been observed also for other QD systems \cite{PhysRevB.93.035311} and can be {\it qualitatively} explained in the SP picture following the arguments of Ref. \cite{doi:10.1063/1.3665634}: the conduction band $e^-$ g-factor is approximately isotropic due to the underlying s-type atomic orbitals; This is not true for the valence band $h^+$ ~\cite{PhysRevB.84.195305,PhysRevB.99.195305,PhysRevLett.96.026804} as the heavy hole (HH) Bloch state has only a projection along the z-direction. Therefore the $h^+$ g-factor is $g_h\approx 0$ along $x$, while along $z$ $g_h \gg 0$ is expected \cite{10.1021/acs.nanolett.6b02715}. The $h^+$ ground state in our QDs has a dominant HH character\cite{Huo:NatPhys}, so we expect a small value of $g_x$ for X, X$^+$, and X$^-$, which is in line with our measurements. The charged states X$^+$ and X$^-$ already allow us to observe a discrepancy from the SP picture. Within that simple model the g-factor of the X$^+$ and X$^-$ is given by $g_{{\rm X}^+}=g_{e,1}+g_{h,2}$ and $g_{{\rm X}^-}=g_{e,2}+g_{h,1}$, respectively, whereby $g_{e(h),1}$ is the single $e^-$ ($h^+$) g-factor in the initial state and $g_{e(h),2}$ the g-factor of the remaining $e^-$ $(h^+)$\cite{PhysRevB.71.075326}. Since the involved $e^-$ and $h^+$ occupy exclusively the ground s levels, we would expect a similar g-factor for $X^+$ and $X^-$, which, however, is not confirmed by our experimental data. Furthermore, X$^+$ does not even show a significant $x$-$z$ anisotropy. We want to point out that X$^+$, X$^-$ and X are linearly polarized under $z$-field, which is not expected from SP theory (see Supplementary Section II) \cite{PhysRevB.65.195315}. In contrast to that, the transitions X$^*$, X$^{+*}$, and X$_{2}^{+*}$-X$_{4}^{+*}$ show a larger $g_{x}$. Interestingly, the g-factor anisotropy is reversed for X$_{2}^{+*}$. As discussed above, in a SP picture, the X$_{2}^{+*}$-X$_{4}^{+*}$ transitions stem from the recombination of a ground-state electron with a ground-state hole in presence of a hole in the first excited state. The pronounced difference in the g-factor anisotropy compared to X$^+$ clearly indicates that the extra hole is by far not simple "spectator" and that its presence and properties (in particular its significant LH content) has profound effects on the response of the resulting exciton to magnetic fields.   
%

For the higher order excitations of the QD we find that X$^*$, X$^{+*}$, X$_4^{+*}$, and X$_3^{+*}$ have larger values of $g_z$ compared to the ground state transitions X, X$^+$, X$^-$. We attribute this to the LH-HH coupling \cite{DURNEV2012797}. In particular, the large $g_z$ value of $X_3^{+*}$ leads to a crossing of one of its components with the $X^+$ states at moderate magnetic fields. For fields above the crossing point such component disappears, possibly because of efficient relaxation to the lower energy $X^+$ state. Furthermore, we observe a coupling between the recombination channels of X$^*$ as expected for a complex including a dark state (see Supplementary Section I).

In contrast to our expectations, the diamagnetic shifts are well fitted by Eq.~(\ref{eq:dia}) also for high magnetic fields. We find $\gamma_{x}<\gamma_{z}$ (except for X$_3^{+*}$). For the neutral X this is qualitatively expected, as the wavefunction is strongly (weakly) confined in $z$ ($x$) directions, leading to a small (large) $\gamma_x$ ($\gamma_z$). For a charged states, a carrier remains in the QD after $e^-$-$h^+$ recombination. Hence, the measured $\gamma$ depends on the localization of the initial as well as the final state. We observe a $\gamma_x$ which is significantly larger for X$_4^{+*}$ and X$_3^{+*}$ compared to the other states and an unexpected high $\gamma_z$ for X$_4^{+*}$. Furthermore, we find a negative diamagnetic coefficient for X$_3^{+*}$. This is known as anomalous diamagnetic shift and was observed for negative trions in InAs/GaAs QDs with weakly confined $e^-$ in the conduction band (see Supplementary Section IV) \cite{PhysRevB.81.113307}. In our QDs, we ascribe the anomalous diamagnetic shift to an initial state which is more localized along $x$ than the remaining hole after recombination.

In order to demonstrate that the obtained results are not a feature of a single QD, we extended our study to several QDs (see Supplementary Section III). It turns out that most of the magneto-optical properties are similar, however, we observe significant differences for the excited trion states X$_2^{+*}-$X$_4^{+*}$, indicating a strong dependence on the structural properties of the QD.

As discussed above, the magnetic properties of the trions contradict the SP model. We now demonstrate that the SP Zeeman Hamiltonian is not even sufficient to describe the magnetic response of a ground state X. We follow Ref. \cite{PhysRevB.84.195305}, where the Hamiltonian of exciton under in-plane magnetic field in the total angular momentum basis is given by:
\begin{equation}
    H_{B}^x=\frac{1}{2}\begin{bmatrix}
\delta_0 & \delta_1 & \epsilon_e & \epsilon_h \\
\delta_1 & \delta_0 & \epsilon_h & \epsilon_e \\
\epsilon_e & \epsilon_h & -\delta_0 & \delta_2 \\
\epsilon_h & \epsilon_e & \delta_2 & -\delta_0
\end{bmatrix},
\label{eq:sp}
\end{equation}
where $\delta_0$ is the splitting between bright and dark exciton states, $\delta_1$ is the exchange splitting between the bright states (fine structure splitting), $\delta_2$ the splitting between the dark states and $\epsilon_{e (h)}=\mu_B B_x g_{e(h),x}$ with $g_{e(h),x}$ the $e^-$ ($h^+$) g-factor in $x$-direction. Due to field-induced bright-dark X coupling, in total four dipole transitions are possible. We label the transitions as X$_{b1,2}$ and X$_{d1,2}$, where X$_{bi}$ (X$_{di}$) are the transitions of the X complex which are bright (dark) at $B=0$ and $i=1,2$ are the respective orthogonally polarized components. The degree of mixing between bright and dark states depends on $B_{x}$ and $g_{e,h}$, whereby the intensity of X$_{b1,2}$ (X$_{d1,2}$) is decreasing (increasing) with increasing field. By calculating the eigenvalues of Eq.~(\ref{eq:sp}) (see Supplementary Section V) we obtain four equations to fit our measurement data. For the fitting we use the measured values for $\delta_0=110$ $\mu$eV and $\delta_1=4.1$ $\mu$eV. 
\begin{figure}[htbp]
	\centering
		\includegraphics[width=85mm]{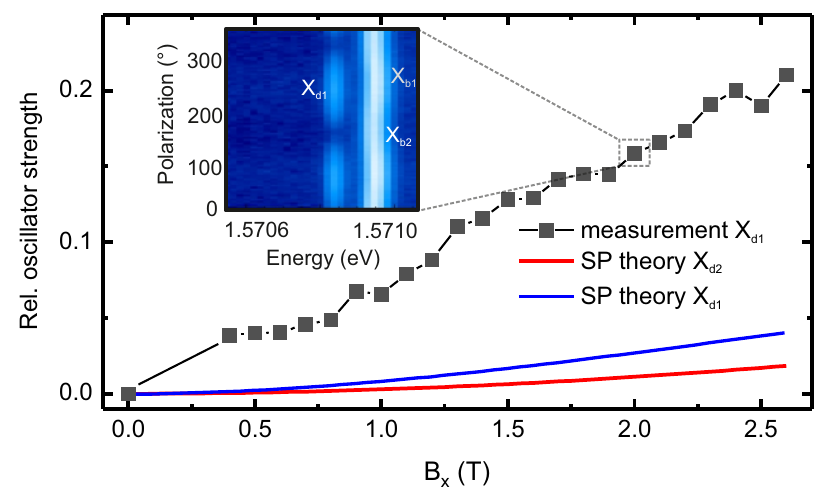}
	\caption{Relative oscillator strength of the dark exciton state versus magnetic field along [110]. The black squares are the measured values for one dark exciton component (X$_{d1}$), the second component (X$_{d2}$) stays dark. The blue (red) curve are the calculated relative oscillator strengths using the $e^-$ and $h^+$ in-plane g-factors extracted by fitting the measurement data via the SP Zeeman Hamiltonian. The inset shows linear-polarization-resolved spectra $B_{x}=2$ T.} 
	\label{fig:fig6}
\end{figure}
The value of $\delta_2$ for dark exciton doublet is not known, as in all measured QDs only the X$_{d1}$ component becomes visible, while X$_{d2}$ stays dark (see inset Fig. \ref{fig:fig6}). However, the dark state splitting is supposed to be small and we assume $\delta_2 \approx 0$. Additionally, we find that within a range of 0-20 $\mu$eV the influence of $\delta_2$ on the mixing is negligible. We obtain from the fit $|g_{e,x}|=0.27$ and $|g_{h,x}|=0.05$. Furthermore, we use the eigenvectors of Eq.~(\ref{eq:sp}) to derive the relative oscillator strength (ROS) $\mathcal{R}_{d1,2}$ of X$_{d1,2}$ (see Supplementary Section V), which determines the coupling between dark and bright states. With the measured values for $\delta_0$ and $\delta_1$ given above, $\delta_2=0$ and the fitted values of $g_{e,x}=0.27$ and $g_{h,x}=-0.05$ we obtain the blue and red curve in Fig. \ref{fig:fig6}. Note that we choose the sign of $g_{e,x}$ and $g_{h,x}$ so that the ROS becomes maximal for one component. As expected, the ROS of X$_{d1,2}$ is increasing by ramping up the magnetic field, due to increased mixing between dark and bright X states.

We also calculate from the measurement data the ROS according to:
\begin{equation}
    \mathcal{R}_{d1}=\frac{I_{{\rm X}_{d1}}}{I_{{\rm X}_{d1}}+I_{{\rm X}_{b1}}},
\label{eq:osc}
\end{equation}
where $I_{{\rm X}_{d1,b1}}$ is the intensity of X$_{d1,b1}$. The result is shown in Fig. \ref{fig:fig6} (black rectangles). Obviously, the measured data do not correspond to the calculated ROS and the coupling between bright and dark states is stronger than expected.

If we use now the ROS equations obtained via the eigenvectors of Eq.~(\ref{eq:sp}) (see Supplementary V) and fit the measured trend in Fig. \ref{fig:fig6} we obtain $g_{h,x}\approx-g_{e,x} \approx 0.52 $, which is different from the result obtained by fitting the energy shift with the eigenvalues. Hence, the SP model is not self-consistent and we can conclude that it is not valid in the weak confinement regime. Interestingly, the SP model yields reasonable results for strongly confining GaAs/AlGaAs QDs \cite{Kunz2013}.

\section{Configuration interaction calculations}

%
%
To gather a deeper insight in the experimental results and support our claim that the SP model is not suitable to extract single-particle g-factors from PL measurements, we perform calculations combining the 8-band ${\bf k}\cdot{\bf p}$ method for the computation of single-particle states and the CI method for the excitonic states confined in our weakly confining QDs. On the one hand this approach allows a realistic treatment of the QD shape and composition~\cite{Birner:07,ArxivKlenovsky:19}, including strain and piezoelectricity up to second order~\cite{Beya-Wakata2011,Klenovsky2018,Aberl:17}. On the other hand it allows an intrinsic treatment of correlation effects, which are included in CI~\cite{Klenovsky2017,ArxivKlenovsky:19} via the excited SP states used to construct the Slater determinants. This is important, because we expect correlation effects between the confined carriers to play a dominant role in the weak confinement regime.

The simulated QD is composed of pure GaAs embedded in an Al$_{0.4}$Ga$_{0.6}$As matrix. Its shape reflects the results of atomic force microscopy measurements on droplet etched nanoholes fabricated under the same growth conditions as the QDs. Additionally, we optimized the structure to match the X emission energy. The final QD shape is such that the QD top is convex while its base is concave (see Supplementary Section VI). 

\begin{figure*}[htbp]
	\centering
		\includegraphics[width=170mm]{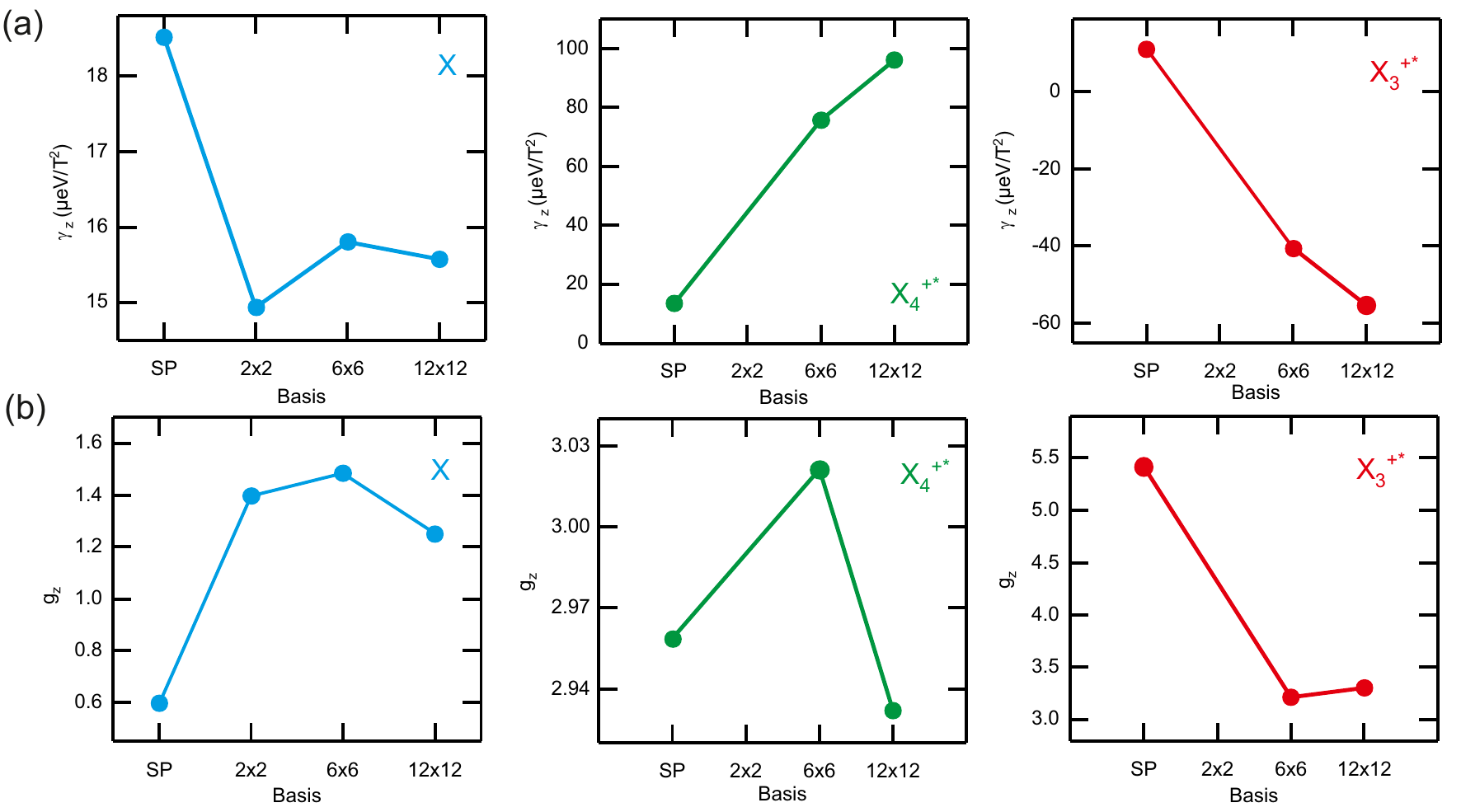}
	\caption{Magnetic parameters of the groundstate exciton (X) and the excited trion states X$_4^{+*}$ and X$_3^{+*}$ confined in a  GaAs QD, calculated with the CI method using single particle states obtained through the ${\bf k}\cdot{\bf p}$ method for a magnetic field along [001]. The results for $\gamma_z$ are given in panel (a) and for $g_z$ in (b) for single particle transition (marked by SP on the horizontal axis) and for two electron and two hole (marked by 2$\times$2), six electron and six hole (6$\times$6), and twelve electron and twelve hole (12$\times$12) CI basis. Note that the effect of correlation increases with the basis size,~i.~e., from SP to ($12 \times 12$) CI basis.}
	\label{fig:fig7}
\end{figure*}

 We start out by calculating the SP recombination of X under a magnetic field along $z$, neglecting Coulomb interaction and correlation. Using Eq.~(\ref{eq:gfactor}) we can extract $\gamma$, $g_{2,z}$, and $g_{z}$ from the computed eigenvalues. The results are presented in Fig. \ref{fig:fig7}. 
The value of $g_z=0.6$ (Fig. \ref{fig:fig7}b left panel) as well as $\gamma=19$ $\mu$eV/T$^2$ (Fig. \ref{fig:fig7}a left panel) show a significant difference from the measurement data. In the next step we include CI in the simulation, where we start with a SP basis including two $e^-$ and two $h^+$ ($2\times2$) states and increase it up to twelve $e^-$ and twelve $h^+$ ($12\times 12$) states. In the CI calculation $\gamma$ ($g_z$) decreases (increases) by almost one quarter (a factor of two) and approaches a value of 15~${\rm \mu eV/T^2}$ (1.25), which is in good agreement with the measurement. We attribute the decrease of the diamagnetic coefficient to the fact that correlation effects lead to a "shrinkage" of the X wavefunction compared to the bare single particle states, so as to improve the overlap to the hole wavefunction.
Furthermore, our calculations point out that the SP model fails to describe the magneto-optical properties in the weak confinement regime and correlation effects cannot be neglected even for the X ground state.  


In addition, we use CI to calculate the g-factors and diamagnetic coefficients for the states provided in Tab.~\ref{tab:tab1}. The results are summarized in the same table (for details on the calculation see Supplementary Section VI), whereby we extend our analysis also to magnetic fields along $x$. The $X^+$ g-factor shows only a small $x$/$z$ anisotropy, which is in good agreement with the experiment. Furthermore, the g-factors of the $X^-$ state are $\approx 2$ times larger than the one of the $X^+$ state, which is not in agreement with the SP description as already discussed in the experimental section.

For the hot trions the CI calculation provides an insight in the origin of large values of $\gamma_z$ and anomalous diamagnetic shift (see Fig. \ref{fig:fig7}, middle and right panels). These phenomena stem from the mixing of the singlet (X$_4^{+*}$) and the triplet (X$_3^{+*}$) trion excited state, an effect that can be traced back to the different magnitude of the electron-hole exchange interaction experienced by each of the two holes constituting hot trion states~\cite{Maialle:07}. The described effect can happen,~e.g., if the excited trion state is spread over a larger region compared to the final hole state. This is the case of the weak confinement which occurs for our QDs in their lateral dimension, hence, the anomalous diamagnetic shift is seen in our case for the Faraday configuration of the applied $B$ field. On the other hand, our QDs are much thinner in the vertical direction and both holes in hot trions are then more strongly confined, with the result that they experience the exchange interaction with electron with equal magnitude. Hence, no large diamagnetic shift is expected for Voigt configuration, exactly as we observe. We stress that the described mixing of singlet and triplet of X$^+$ for $z$-direction is a purely multi-particle effect (not describable on SP level) and, moreover, occurs because of the small energy separation between X$_4^{+*}$ and X$_3^{+ *}$ as seen in our experiments and calculations, where that amounts to $\approx 100\,\mu{\rm eV}$ (see Supplementary Section VI). On the other hand the value of $g_z$ is mostly determined by the Zeeman splitting of the final SP hole states that are subtracted from trions (see also Supplementary Section VI).

In general, the simulation results are qualitatively in good agreement with the measurement data. We partly attribute the mismatch in the absolute values between measurement and calculation to differences in the shape and size of the measured and simulated QD. Nevertheless, the CI also shows some deviations from the experiment: Due to the strong confinement along $z$ we would expect similar values for $\gamma_x$ for all states, which is in line with the measurement. However, we obtain for X$_3^{+*}$ a  $\gamma_x=2$ $\mu eV/T^2$ from the calculations, which is about 3 times smaller than expected. Here the $\gamma_x$ value directly obtained from the SP states obtained by ${\bf k}\cdot{\bf p}$ is closer to the experimental one (see Supplementary Figure 10). Finally we note that, the calculated binding energy of X$^+$ with respect to the X is only 600 $\mu$eV, whereby all measured QDs show a value of $\approx 2.5$ meV. The reason for this deviation is not fully understood yet, but since the problem was observed in independent CI calculations \cite{PhysRevB.80.085309}, we speculate that it may stem from an intrinsic limitation of the CI method.

\section{Conclusion}

In conclusion, we have presented a comprehensive analysis of the optical transitions in a GaAs QD under above bandgap excitation. The performed measurements allow us to determine the charge complexes forming the excited states in a QD. Furthermore, we provide an analysis of the g-factors and diamagnetic coefficients of the excited complexes in our GaAs/AlGaAs QDs obtained by droplet etching. On this basis we are able to experimentally prove that the SP Zeemann Hamiltonian \cite{PhysRevB.65.195315} cannot be used to reliably extract single-particle g-factors from measurements of optical transitions in weakly confining QDs. CI calculations clearly show that interactions between the confined carriers --- such as correlation effects --- significantly influence the magneto-optical properties. The model calculations are able to quantitatively reproduce most of the observed values of diamagnetic shifts and g-factors not only for the neutral excitons, but also for some of the charged complexes observed experimentally. Some significant differences between experiment and calculation results are still present, which deserve further consideration in the future.

\section*{Acknowledgement}
We thank L. Vukusic, G. Katsaros, F. Binder and B. Swolo for fruitful discussions and technical assistance.

This work was supported by the austrian science fund (FWF): P29603 , the Linz Institute of Technology (LIT) and the LIT Secure and Correct Systems Lab, financed by the state of Upper Austria. 

P.K and D.C received national funding from project CEITEC 2020 (LQ1601) with financial support from the Ministry of Education, Youth and Sports of the Czech Republic under the National Sustainability Programme II and the funding from European Union's Horizon 2020 (2014-2020) research and innovation framework program under grant agreement No 731473. P.K. was (partially) funded by project EMPIR 17FUN06 Siqust. This project has received funding from the EMPIR programme co-financed by the Participating States and from the European Union’s Horizon 2020 research and innovation programme.
\newpage

\bibliography{references}

\begin{thebibliography}{54}%
\makeatletter
\providecommand \@ifxundefined [1]{%
 \@ifx{#1\undefined}
}%
\providecommand \@ifnum [1]{%
 \ifnum #1\expandafter \@firstoftwo
 \else \expandafter \@secondoftwo
 \fi
}%
\providecommand \@ifx [1]{%
 \ifx #1\expandafter \@firstoftwo
 \else \expandafter \@secondoftwo
 \fi
}%
\providecommand \natexlab [1]{#1}%
\providecommand \enquote  [1]{``#1''}%
\providecommand \bibnamefont  [1]{#1}%
\providecommand \bibfnamefont [1]{#1}%
\providecommand \citenamefont [1]{#1}%
\providecommand \href@noop [0]{\@secondoftwo}%
\providecommand \href [0]{\begingroup \@sanitize@url \@href}%
\providecommand \@href[1]{\@@startlink{#1}\@@href}%
\providecommand \@@href[1]{\endgroup#1\@@endlink}%
\providecommand \@sanitize@url [0]{\catcode `\\12\catcode `\$12\catcode
  `\&12\catcode `\#12\catcode `\^12\catcode `\_12\catcode `\%12\relax}%
\providecommand \@@startlink[1]{}%
\providecommand \@@endlink[0]{}%
\providecommand \url  [0]{\begingroup\@sanitize@url \@url }%
\providecommand \@url [1]{\endgroup\@href {#1}{\urlprefix }}%
\providecommand \urlprefix  [0]{URL }%
\providecommand \Eprint [0]{\href }%
\providecommand \doibase [0]{http://dx.doi.org/}%
\providecommand \selectlanguage [0]{\@gobble}%
\providecommand \bibinfo  [0]{\@secondoftwo}%
\providecommand \bibfield  [0]{\@secondoftwo}%
\providecommand \translation [1]{[#1]}%
\providecommand \BibitemOpen [0]{}%
\providecommand \bibitemStop [0]{}%
\providecommand \bibitemNoStop [0]{.\EOS\space}%
\providecommand \EOS [0]{\spacefactor3000\relax}%
\providecommand \BibitemShut  [1]{\csname bibitem#1\endcsname}%
\let\auto@bib@innerbib\@empty
\bibitem [{\citenamefont {Wang}\ \emph {et~al.}(2007)\citenamefont {Wang},
  \citenamefont {Liang}, \citenamefont {Sablon},\ and\ \citenamefont
  {Salamo}}]{Wang2007}%
  \BibitemOpen
  \bibfield  {author} {\bibinfo {author} {\bibfnamefont {Z.~M.}\ \bibnamefont
  {Wang}}, \bibinfo {author} {\bibfnamefont {B.~L.}\ \bibnamefont {Liang}},
  \bibinfo {author} {\bibfnamefont {K.~A.}\ \bibnamefont {Sablon}}, \ and\
  \bibinfo {author} {\bibfnamefont {G.~J.}\ \bibnamefont {Salamo}},\ }\href
  {\doibase 10.1063/1.2713745} {\bibfield  {journal} {\bibinfo  {journal}
  {Applied Physics Letters}\ }\textbf {\bibinfo {volume} {90}},\ \bibinfo
  {pages} {113120} (\bibinfo {year} {2007})}\BibitemShut {NoStop}%
\bibitem [{\citenamefont {Heyn}\ \emph {et~al.}(2009)\citenamefont {Heyn},
  \citenamefont {Stemmann}, \citenamefont {K{\"{o}}ppen}, \citenamefont
  {Strelow}, \citenamefont {Kipp}, \citenamefont {Grave}, \citenamefont
  {Mendach},\ and\ \citenamefont {Hansen}}]{Heyn2009}%
  \BibitemOpen
  \bibfield  {author} {\bibinfo {author} {\bibfnamefont {C.}~\bibnamefont
  {Heyn}}, \bibinfo {author} {\bibfnamefont {A.}~\bibnamefont {Stemmann}},
  \bibinfo {author} {\bibfnamefont {T.}~\bibnamefont {K{\"{o}}ppen}}, \bibinfo
  {author} {\bibfnamefont {C.}~\bibnamefont {Strelow}}, \bibinfo {author}
  {\bibfnamefont {T.}~\bibnamefont {Kipp}}, \bibinfo {author} {\bibfnamefont
  {M.}~\bibnamefont {Grave}}, \bibinfo {author} {\bibfnamefont
  {S.}~\bibnamefont {Mendach}}, \ and\ \bibinfo {author} {\bibfnamefont
  {W.}~\bibnamefont {Hansen}},\ }\href@noop {} {\bibfield  {journal} {\bibinfo
  {journal} {Applied Physics Letters}\ }\textbf {\bibinfo {volume} {94}},\
  \bibinfo {pages} {18} (\bibinfo {year} {2009})}\BibitemShut {NoStop}%
\bibitem [{\citenamefont {Huo}\ \emph {et~al.}(2013)\citenamefont {Huo},
  \citenamefont {Rastelli},\ and\ \citenamefont {Schmidt}}]{Huo:APL2013}%
  \BibitemOpen
  \bibfield  {author} {\bibinfo {author} {\bibfnamefont {Y.~H.}\ \bibnamefont
  {Huo}}, \bibinfo {author} {\bibfnamefont {A.}~\bibnamefont {Rastelli}}, \
  and\ \bibinfo {author} {\bibfnamefont {O.~G.}\ \bibnamefont {Schmidt}},\
  }\href@noop {} {\bibfield  {journal} {\bibinfo  {journal} {Applied Physics
  Letters}\ }\textbf {\bibinfo {volume} {102}},\ \bibinfo {pages} {152105}
  (\bibinfo {year} {2013})}\BibitemShut {NoStop}%
\bibitem [{\citenamefont {Schweickert}\ \emph {et~al.}(2018)\citenamefont
  {Schweickert}, \citenamefont {Jöns}, \citenamefont {Zeuner}, \citenamefont
  {Covre~da Silva}, \citenamefont {Huang}, \citenamefont {Lettner},
  \citenamefont {Reindl}, \citenamefont {Zichi}, \citenamefont {Trotta},
  \citenamefont {Rastelli},\ and\ \citenamefont
  {Zwiller}}]{doi:10.1063/1.5020038}%
  \BibitemOpen
  \bibfield  {author} {\bibinfo {author} {\bibfnamefont {L.}~\bibnamefont
  {Schweickert}}, \bibinfo {author} {\bibfnamefont {K.~D.}\ \bibnamefont
  {Jöns}}, \bibinfo {author} {\bibfnamefont {K.~D.}\ \bibnamefont {Zeuner}},
  \bibinfo {author} {\bibfnamefont {S.~F.}\ \bibnamefont {Covre~da Silva}},
  \bibinfo {author} {\bibfnamefont {H.}~\bibnamefont {Huang}}, \bibinfo
  {author} {\bibfnamefont {T.}~\bibnamefont {Lettner}}, \bibinfo {author}
  {\bibfnamefont {M.}~\bibnamefont {Reindl}}, \bibinfo {author} {\bibfnamefont
  {J.}~\bibnamefont {Zichi}}, \bibinfo {author} {\bibfnamefont
  {R.}~\bibnamefont {Trotta}}, \bibinfo {author} {\bibfnamefont
  {A.}~\bibnamefont {Rastelli}}, \ and\ \bibinfo {author} {\bibfnamefont
  {V.}~\bibnamefont {Zwiller}},\ }\href {\doibase 10.1063/1.5020038} {\bibfield
   {journal} {\bibinfo  {journal} {Applied Physics Letters}\ }\textbf {\bibinfo
  {volume} {112}},\ \bibinfo {pages} {093106} (\bibinfo {year} {2018})},\
  \Eprint {http://arxiv.org/abs/https://doi.org/10.1063/1.5020038}
  {https://doi.org/10.1063/1.5020038} \BibitemShut {NoStop}%
\bibitem [{\citenamefont {Huber}\ \emph {et~al.}(2017)\citenamefont {Huber},
  \citenamefont {Reindl}, \citenamefont {Huo}, \citenamefont {Huang},
  \citenamefont {Wildmann}, \citenamefont {Schmidt}, \citenamefont {Rastelli},\
  and\ \citenamefont {Trotta}}]{Huber2017}%
  \BibitemOpen
  \bibfield  {author} {\bibinfo {author} {\bibfnamefont {D.}~\bibnamefont
  {Huber}}, \bibinfo {author} {\bibfnamefont {M.}~\bibnamefont {Reindl}},
  \bibinfo {author} {\bibfnamefont {Y.}~\bibnamefont {Huo}}, \bibinfo {author}
  {\bibfnamefont {H.}~\bibnamefont {Huang}}, \bibinfo {author} {\bibfnamefont
  {J.~S.}\ \bibnamefont {Wildmann}}, \bibinfo {author} {\bibfnamefont {O.~G.}\
  \bibnamefont {Schmidt}}, \bibinfo {author} {\bibfnamefont {A.}~\bibnamefont
  {Rastelli}}, \ and\ \bibinfo {author} {\bibfnamefont {R.}~\bibnamefont
  {Trotta}},\ }\href@noop {} {\bibfield  {journal} {\bibinfo  {journal} {Nature
  Communications}\ }\textbf {\bibinfo {volume} {8}},\ \bibinfo {pages} {15506}
  (\bibinfo {year} {2017})}\BibitemShut {NoStop}%
\bibitem [{\citenamefont {Reindl}\ \emph {et~al.}(2019)\citenamefont {Reindl},
  \citenamefont {Weber}, \citenamefont {Huber}, \citenamefont {Schimpf},
  \citenamefont {Covre~da Silva}, \citenamefont {Portalupi}, \citenamefont
  {Trotta}, \citenamefont {Michler},\ and\ \citenamefont
  {Rastelli}}]{Reindl2019}%
  \BibitemOpen
  \bibfield  {author} {\bibinfo {author} {\bibfnamefont {M.}~\bibnamefont
  {Reindl}}, \bibinfo {author} {\bibfnamefont {J.~H.}\ \bibnamefont {Weber}},
  \bibinfo {author} {\bibfnamefont {D.}~\bibnamefont {Huber}}, \bibinfo
  {author} {\bibfnamefont {C.}~\bibnamefont {Schimpf}}, \bibinfo {author}
  {\bibfnamefont {S.~F.}\ \bibnamefont {Covre~da Silva}}, \bibinfo {author}
  {\bibfnamefont {S.~L.}\ \bibnamefont {Portalupi}}, \bibinfo {author}
  {\bibfnamefont {R.}~\bibnamefont {Trotta}}, \bibinfo {author} {\bibfnamefont
  {P.}~\bibnamefont {Michler}}, \ and\ \bibinfo {author} {\bibfnamefont
  {A.}~\bibnamefont {Rastelli}},\ }\href@noop {} {\bibfield  {journal}
  {\bibinfo  {journal} {Preprint at: arXiv:1901.11251}\ } (\bibinfo {year}
  {2019})}\BibitemShut {NoStop}%
\bibitem [{\citenamefont {Sch\"oll}\ \emph {et~al.}(2019)\citenamefont
  {Sch\"oll}, \citenamefont {Hanschke}, \citenamefont {Schweickert},
  \citenamefont {Zeuner}, \citenamefont {Reindl}, \citenamefont {Covre~da
  Silva}, \citenamefont {Lettner}, \citenamefont {Trotta}, \citenamefont
  {Finley}, \citenamefont {Müller}, \citenamefont {Rastelli}, \citenamefont
  {Zwiller},\ and\ \citenamefont {Jöns}}]{doi:10.1021/acs.nanolett.8b05132}%
  \BibitemOpen
  \bibfield  {author} {\bibinfo {author} {\bibfnamefont {E.}~\bibnamefont
  {Sch\"oll}}, \bibinfo {author} {\bibfnamefont {L.}~\bibnamefont {Hanschke}},
  \bibinfo {author} {\bibfnamefont {L.}~\bibnamefont {Schweickert}}, \bibinfo
  {author} {\bibfnamefont {K.~D.}\ \bibnamefont {Zeuner}}, \bibinfo {author}
  {\bibfnamefont {M.}~\bibnamefont {Reindl}}, \bibinfo {author} {\bibfnamefont
  {S.~F.}\ \bibnamefont {Covre~da Silva}}, \bibinfo {author} {\bibfnamefont
  {T.}~\bibnamefont {Lettner}}, \bibinfo {author} {\bibfnamefont
  {R.}~\bibnamefont {Trotta}}, \bibinfo {author} {\bibfnamefont {J.~J.}\
  \bibnamefont {Finley}}, \bibinfo {author} {\bibfnamefont {K.}~\bibnamefont
  {Müller}}, \bibinfo {author} {\bibfnamefont {A.}~\bibnamefont {Rastelli}},
  \bibinfo {author} {\bibfnamefont {V.}~\bibnamefont {Zwiller}}, \ and\
  \bibinfo {author} {\bibfnamefont {K.~D.}\ \bibnamefont {Jöns}},\ }\href
  {\doibase 10.1021/acs.nanolett.8b05132} {\bibfield  {journal} {\bibinfo
  {journal} {Nano Letters}\ }\textbf {\bibinfo {volume} {19}},\ \bibinfo
  {pages} {2404} (\bibinfo {year} {2019})},\ \bibinfo {note} {pMID: 30862165},\
  \Eprint {http://arxiv.org/abs/https://doi.org/10.1021/acs.nanolett.8b05132}
  {https://doi.org/10.1021/acs.nanolett.8b05132} \BibitemShut {NoStop}%
\bibitem [{\citenamefont {Liu}\ \emph {et~al.}(2019)\citenamefont {Liu},
  \citenamefont {Su}, \citenamefont {Wei}, \citenamefont {Yao}, \citenamefont
  {Filipe}, \citenamefont {Yu}, \citenamefont {Iles-smith}, \citenamefont
  {Srinivasan}, \citenamefont {Rastelli}, \citenamefont {Li},\ and\
  \citenamefont {Wang}}]{Liu2019}%
  \BibitemOpen
  \bibfield  {author} {\bibinfo {author} {\bibfnamefont {J.}~\bibnamefont
  {Liu}}, \bibinfo {author} {\bibfnamefont {R.}~\bibnamefont {Su}}, \bibinfo
  {author} {\bibfnamefont {Y.}~\bibnamefont {Wei}}, \bibinfo {author}
  {\bibfnamefont {B.}~\bibnamefont {Yao}}, \bibinfo {author} {\bibfnamefont
  {S.}~\bibnamefont {Filipe}}, \bibinfo {author} {\bibfnamefont
  {Y.}~\bibnamefont {Yu}}, \bibinfo {author} {\bibfnamefont {J.}~\bibnamefont
  {Iles-smith}}, \bibinfo {author} {\bibfnamefont {K.}~\bibnamefont
  {Srinivasan}}, \bibinfo {author} {\bibfnamefont {A.}~\bibnamefont
  {Rastelli}}, \bibinfo {author} {\bibfnamefont {J.}~\bibnamefont {Li}}, \ and\
  \bibinfo {author} {\bibfnamefont {X.}~\bibnamefont {Wang}},\ }\href {\doibase
  10.1038/s41565-019-0435-9} {\bibfield  {journal} {\bibinfo  {journal} {Nature
  Nanotechnology}\ ,\ \bibinfo {pages} {1748}} (\bibinfo {year}
  {2019})}\BibitemShut {NoStop}%
\bibitem [{\citenamefont {Keil}\ \emph {et~al.}(2017)\citenamefont {Keil},
  \citenamefont {Zopf}, \citenamefont {Chen}, \citenamefont {H{\"{o}}fer},
  \citenamefont {Zhang}, \citenamefont {Ding},\ and\ \citenamefont
  {Schmidt}}]{Keil2017}%
  \BibitemOpen
  \bibfield  {author} {\bibinfo {author} {\bibfnamefont {R.}~\bibnamefont
  {Keil}}, \bibinfo {author} {\bibfnamefont {M.}~\bibnamefont {Zopf}}, \bibinfo
  {author} {\bibfnamefont {Y.}~\bibnamefont {Chen}}, \bibinfo {author}
  {\bibfnamefont {B.}~\bibnamefont {H{\"{o}}fer}}, \bibinfo {author}
  {\bibfnamefont {J.}~\bibnamefont {Zhang}}, \bibinfo {author} {\bibfnamefont
  {F.}~\bibnamefont {Ding}}, \ and\ \bibinfo {author} {\bibfnamefont {O.~G.}\
  \bibnamefont {Schmidt}},\ }\href@noop {} {\bibfield  {journal} {\bibinfo
  {journal} {Nature Communications}\ }\textbf {\bibinfo {volume} {8}} (\bibinfo
  {year} {2017})}\BibitemShut {NoStop}%
\bibitem [{\citenamefont {Huber}\ \emph
  {et~al.}(2018{\natexlab{a}})\citenamefont {Huber}, \citenamefont {Reindl},
  \citenamefont {Covre~da Silva}, \citenamefont {Schimpf}, \citenamefont
  {Mart\'{\i}n-S\'anchez}, \citenamefont {Huang}, \citenamefont {Piredda},
  \citenamefont {Edlinger}, \citenamefont {Rastelli},\ and\ \citenamefont
  {Trotta}}]{Huber2018}%
  \BibitemOpen
  \bibfield  {author} {\bibinfo {author} {\bibfnamefont {D.}~\bibnamefont
  {Huber}}, \bibinfo {author} {\bibfnamefont {M.}~\bibnamefont {Reindl}},
  \bibinfo {author} {\bibfnamefont {S.~F.}\ \bibnamefont {Covre~da Silva}},
  \bibinfo {author} {\bibfnamefont {C.}~\bibnamefont {Schimpf}}, \bibinfo
  {author} {\bibfnamefont {J.}~\bibnamefont {Mart\'{\i}n-S\'anchez}}, \bibinfo
  {author} {\bibfnamefont {H.}~\bibnamefont {Huang}}, \bibinfo {author}
  {\bibfnamefont {G.}~\bibnamefont {Piredda}}, \bibinfo {author} {\bibfnamefont
  {J.}~\bibnamefont {Edlinger}}, \bibinfo {author} {\bibfnamefont
  {A.}~\bibnamefont {Rastelli}}, \ and\ \bibinfo {author} {\bibfnamefont
  {R.}~\bibnamefont {Trotta}},\ }\href@noop {} {\bibfield  {journal} {\bibinfo
  {journal} {Phys. Rev. Lett.}\ }\textbf {\bibinfo {volume} {121}},\ \bibinfo
  {pages} {033902} (\bibinfo {year} {2018}{\natexlab{a}})}\BibitemShut
  {NoStop}%
\bibitem [{\citenamefont {Gurioli}\ \emph {et~al.}(2019)\citenamefont
  {Gurioli}, \citenamefont {Wang}, \citenamefont {Rastelli}, \citenamefont
  {Kuroda},\ and\ \citenamefont {Sanguinetti}}]{Gurioli2019}%
  \BibitemOpen
  \bibfield  {author} {\bibinfo {author} {\bibfnamefont {M.}~\bibnamefont
  {Gurioli}}, \bibinfo {author} {\bibfnamefont {Z.}~\bibnamefont {Wang}},
  \bibinfo {author} {\bibfnamefont {A.}~\bibnamefont {Rastelli}}, \bibinfo
  {author} {\bibfnamefont {T.}~\bibnamefont {Kuroda}}, \ and\ \bibinfo {author}
  {\bibfnamefont {S.}~\bibnamefont {Sanguinetti}},\ }\href@noop {} {\bibfield
  {journal} {\bibinfo  {journal} {Nature Materials}\ }\textbf {\bibinfo
  {volume} {18}},\ \bibinfo {pages} {799–810} (\bibinfo {year}
  {2019})}\BibitemShut {NoStop}%
\bibitem [{\citenamefont {Stobbe}\ \emph {et~al.}(2010)\citenamefont {Stobbe},
  \citenamefont {Schlereth}, \citenamefont {H\"ofling}, \citenamefont
  {Forchel}, \citenamefont {Hvam},\ and\ \citenamefont
  {Lodahl}}]{PhysRevB.82.233302}%
  \BibitemOpen
  \bibfield  {author} {\bibinfo {author} {\bibfnamefont {S.}~\bibnamefont
  {Stobbe}}, \bibinfo {author} {\bibfnamefont {T.~W.}\ \bibnamefont
  {Schlereth}}, \bibinfo {author} {\bibfnamefont {S.}~\bibnamefont
  {H\"ofling}}, \bibinfo {author} {\bibfnamefont {A.}~\bibnamefont {Forchel}},
  \bibinfo {author} {\bibfnamefont {J.~M.}\ \bibnamefont {Hvam}}, \ and\
  \bibinfo {author} {\bibfnamefont {P.}~\bibnamefont {Lodahl}},\ }\href
  {\doibase 10.1103/PhysRevB.82.233302} {\bibfield  {journal} {\bibinfo
  {journal} {Phys. Rev. B}\ }\textbf {\bibinfo {volume} {82}},\ \bibinfo
  {pages} {233302} (\bibinfo {year} {2010})}\BibitemShut {NoStop}%
\bibitem [{\citenamefont {Huber}\ \emph
  {et~al.}(2018{\natexlab{b}})\citenamefont {Huber}, \citenamefont {Reindl},
  \citenamefont {Aberl}, \citenamefont {Rastelli},\ and\ \citenamefont
  {Trotta}}]{Huber:rev2018}%
  \BibitemOpen
  \bibfield  {author} {\bibinfo {author} {\bibfnamefont {D.}~\bibnamefont
  {Huber}}, \bibinfo {author} {\bibfnamefont {M.}~\bibnamefont {Reindl}},
  \bibinfo {author} {\bibfnamefont {J.}~\bibnamefont {Aberl}}, \bibinfo
  {author} {\bibfnamefont {A.}~\bibnamefont {Rastelli}}, \ and\ \bibinfo
  {author} {\bibfnamefont {R.}~\bibnamefont {Trotta}},\ }\href@noop {}
  {\bibfield  {journal} {\bibinfo  {journal} {Journal of Optics}\ }\textbf
  {\bibinfo {volume} {20}},\ \bibinfo {pages} {073002} (\bibinfo {year}
  {2018}{\natexlab{b}})}\BibitemShut {NoStop}%
\bibitem [{\citenamefont {Peter}\ \emph {et~al.}(2005)\citenamefont {Peter},
  \citenamefont {Senellart}, \citenamefont {Martrou}, \citenamefont
  {Lema\^{\i}tre}, \citenamefont {Hours}, \citenamefont {G\'erard},\ and\
  \citenamefont {Bloch}}]{PhysRevLett.95.067401}%
  \BibitemOpen
  \bibfield  {author} {\bibinfo {author} {\bibfnamefont {E.}~\bibnamefont
  {Peter}}, \bibinfo {author} {\bibfnamefont {P.}~\bibnamefont {Senellart}},
  \bibinfo {author} {\bibfnamefont {D.}~\bibnamefont {Martrou}}, \bibinfo
  {author} {\bibfnamefont {A.}~\bibnamefont {Lema\^{\i}tre}}, \bibinfo {author}
  {\bibfnamefont {J.}~\bibnamefont {Hours}}, \bibinfo {author} {\bibfnamefont
  {J.~M.}\ \bibnamefont {G\'erard}}, \ and\ \bibinfo {author} {\bibfnamefont
  {J.}~\bibnamefont {Bloch}},\ }\href {\doibase 10.1103/PhysRevLett.95.067401}
  {\bibfield  {journal} {\bibinfo  {journal} {Phys. Rev. Lett.}\ }\textbf
  {\bibinfo {volume} {95}},\ \bibinfo {pages} {067401} (\bibinfo {year}
  {2005})}\BibitemShut {NoStop}%
\bibitem [{\citenamefont {Loss}\ and\ \citenamefont
  {DiVincenzo}(1998)}]{PhysRevA.57.120}%
  \BibitemOpen
  \bibfield  {author} {\bibinfo {author} {\bibfnamefont {D.}~\bibnamefont
  {Loss}}\ and\ \bibinfo {author} {\bibfnamefont {D.~P.}\ \bibnamefont
  {DiVincenzo}},\ }\href {\doibase 10.1103/PhysRevA.57.120} {\bibfield
  {journal} {\bibinfo  {journal} {Phys. Rev. A}\ }\textbf {\bibinfo {volume}
  {57}},\ \bibinfo {pages} {120} (\bibinfo {year} {1998})}\BibitemShut
  {NoStop}%
\bibitem [{\citenamefont {Schwartz}\ \emph {et~al.}(2015)\citenamefont
  {Schwartz}, \citenamefont {Schmidgall}, \citenamefont {Gantz}, \citenamefont
  {Cogan}, \citenamefont {Bordo}, \citenamefont {Don}, \citenamefont
  {Zielinski},\ and\ \citenamefont {Gershoni}}]{PhysRevX.5.011009}%
  \BibitemOpen
  \bibfield  {author} {\bibinfo {author} {\bibfnamefont {I.}~\bibnamefont
  {Schwartz}}, \bibinfo {author} {\bibfnamefont {E.~R.}\ \bibnamefont
  {Schmidgall}}, \bibinfo {author} {\bibfnamefont {L.}~\bibnamefont {Gantz}},
  \bibinfo {author} {\bibfnamefont {D.}~\bibnamefont {Cogan}}, \bibinfo
  {author} {\bibfnamefont {E.}~\bibnamefont {Bordo}}, \bibinfo {author}
  {\bibfnamefont {Y.}~\bibnamefont {Don}}, \bibinfo {author} {\bibfnamefont
  {M.}~\bibnamefont {Zielinski}}, \ and\ \bibinfo {author} {\bibfnamefont
  {D.}~\bibnamefont {Gershoni}},\ }\href {\doibase 10.1103/PhysRevX.5.011009}
  {\bibfield  {journal} {\bibinfo  {journal} {Phys. Rev. X}\ }\textbf {\bibinfo
  {volume} {5}},\ \bibinfo {pages} {011009} (\bibinfo {year}
  {2015})}\BibitemShut {NoStop}%
\bibitem [{\citenamefont {Gaudreau}\ \emph {et~al.}(2017)\citenamefont
  {Gaudreau}, \citenamefont {Bogan}, \citenamefont {Korkusinski}, \citenamefont
  {Studenikin}, \citenamefont {Austing},\ and\ \citenamefont
  {Sachrajda}}]{Gaudreau_2017}%
  \BibitemOpen
  \bibfield  {author} {\bibinfo {author} {\bibfnamefont {L.}~\bibnamefont
  {Gaudreau}}, \bibinfo {author} {\bibfnamefont {A.}~\bibnamefont {Bogan}},
  \bibinfo {author} {\bibfnamefont {M.}~\bibnamefont {Korkusinski}}, \bibinfo
  {author} {\bibfnamefont {S.}~\bibnamefont {Studenikin}}, \bibinfo {author}
  {\bibfnamefont {D.~G.}\ \bibnamefont {Austing}}, \ and\ \bibinfo {author}
  {\bibfnamefont {A.~S.}\ \bibnamefont {Sachrajda}},\ }\href {\doibase
  10.1088/1361-6641/aa788d} {\bibfield  {journal} {\bibinfo  {journal}
  {Semiconductor Science and Technology}\ }\textbf {\bibinfo {volume} {32}},\
  \bibinfo {pages} {093001} (\bibinfo {year} {2017})}\BibitemShut {NoStop}%
\bibitem [{\citenamefont {Tholen}\ \emph {et~al.}(2019)\citenamefont {Tholen},
  \citenamefont {Wildmann}, \citenamefont {Rastelli}, \citenamefont {Trotta},
  \citenamefont {Pryor}, \citenamefont {Zallo}, \citenamefont {Schmidt},
  \citenamefont {Koenraad},\ and\ \citenamefont {Silov}}]{PhysRevB.99.195305}%
  \BibitemOpen
  \bibfield  {author} {\bibinfo {author} {\bibfnamefont {H.~M. G.~A.}\
  \bibnamefont {Tholen}}, \bibinfo {author} {\bibfnamefont {J.~S.}\
  \bibnamefont {Wildmann}}, \bibinfo {author} {\bibfnamefont {A.}~\bibnamefont
  {Rastelli}}, \bibinfo {author} {\bibfnamefont {R.}~\bibnamefont {Trotta}},
  \bibinfo {author} {\bibfnamefont {C.~E.}\ \bibnamefont {Pryor}}, \bibinfo
  {author} {\bibfnamefont {E.}~\bibnamefont {Zallo}}, \bibinfo {author}
  {\bibfnamefont {O.~G.}\ \bibnamefont {Schmidt}}, \bibinfo {author}
  {\bibfnamefont {P.~M.}\ \bibnamefont {Koenraad}}, \ and\ \bibinfo {author}
  {\bibfnamefont {A.~Y.}\ \bibnamefont {Silov}},\ }\href {\doibase
  10.1103/PhysRevB.99.195305} {\bibfield  {journal} {\bibinfo  {journal} {Phys.
  Rev. B}\ }\textbf {\bibinfo {volume} {99}},\ \bibinfo {pages} {195305}
  (\bibinfo {year} {2019})}\BibitemShut {NoStop}%
\bibitem [{\citenamefont {Bennett}\ \emph {et~al.}(2013)\citenamefont
  {Bennett}, \citenamefont {Pooley}, \citenamefont {Cao}, \citenamefont
  {Sklöd}, \citenamefont {Farrer}, \citenamefont {Ritchie},\ and\
  \citenamefont {Shields}}]{Bennett2013}%
  \BibitemOpen
  \bibfield  {author} {\bibinfo {author} {\bibfnamefont {A.~J.}\ \bibnamefont
  {Bennett}}, \bibinfo {author} {\bibfnamefont {M.~A.}\ \bibnamefont {Pooley}},
  \bibinfo {author} {\bibfnamefont {Y.}~\bibnamefont {Cao}}, \bibinfo {author}
  {\bibfnamefont {N.}~\bibnamefont {Sklöd}}, \bibinfo {author} {\bibfnamefont
  {I.}~\bibnamefont {Farrer}}, \bibinfo {author} {\bibfnamefont {D.~A.}\
  \bibnamefont {Ritchie}}, \ and\ \bibinfo {author} {\bibfnamefont {A.~J.}\
  \bibnamefont {Shields}},\ }\href {\doibase 10.1038/ncomms2519} {\bibfield
  {journal} {\bibinfo  {journal} {Nature Comm.}\ }\textbf {\bibinfo {volume}
  {4}},\ \bibinfo {pages} {1522} (\bibinfo {year} {2013})}\BibitemShut
  {NoStop}%
\bibitem [{\citenamefont {Medeiros-Ribeiro}\ \emph {et~al.}(2003)\citenamefont
  {Medeiros-Ribeiro}, \citenamefont {Ribeiro},\ and\ \citenamefont
  {Westfahl~Jr.}}]{Medeiros-Ribeiro2003}%
  \BibitemOpen
  \bibfield  {author} {\bibinfo {author} {\bibfnamefont {G.}~\bibnamefont
  {Medeiros-Ribeiro}}, \bibinfo {author} {\bibfnamefont {E.}~\bibnamefont
  {Ribeiro}}, \ and\ \bibinfo {author} {\bibfnamefont {H.}~\bibnamefont
  {Westfahl~Jr.}},\ }\href {\doibase 10.1007/s00339-003-2241-2} {\bibfield
  {journal} {\bibinfo  {journal} {Applied Physics A}\ }\textbf {\bibinfo
  {volume} {77}},\ \bibinfo {pages} {725} (\bibinfo {year} {2003})}\BibitemShut
  {NoStop}%
\bibitem [{\citenamefont {Bayer}\ \emph {et~al.}(2002)\citenamefont {Bayer},
  \citenamefont {Ortner}, \citenamefont {Stern}, \citenamefont {Kuther},
  \citenamefont {Gorbunov}, \citenamefont {Forchel}, \citenamefont {Hawrylak},
  \citenamefont {Fafard}, \citenamefont {Hinzer}, \citenamefont {Reinecke},
  \citenamefont {Walck}, \citenamefont {Reithmaier}, \citenamefont {Klopf},\
  and\ \citenamefont {Sch\"afer}}]{PhysRevB.65.195315}%
  \BibitemOpen
  \bibfield  {author} {\bibinfo {author} {\bibfnamefont {M.}~\bibnamefont
  {Bayer}}, \bibinfo {author} {\bibfnamefont {G.}~\bibnamefont {Ortner}},
  \bibinfo {author} {\bibfnamefont {O.}~\bibnamefont {Stern}}, \bibinfo
  {author} {\bibfnamefont {A.}~\bibnamefont {Kuther}}, \bibinfo {author}
  {\bibfnamefont {A.~A.}\ \bibnamefont {Gorbunov}}, \bibinfo {author}
  {\bibfnamefont {A.}~\bibnamefont {Forchel}}, \bibinfo {author} {\bibfnamefont
  {P.}~\bibnamefont {Hawrylak}}, \bibinfo {author} {\bibfnamefont
  {S.}~\bibnamefont {Fafard}}, \bibinfo {author} {\bibfnamefont
  {K.}~\bibnamefont {Hinzer}}, \bibinfo {author} {\bibfnamefont {T.~L.}\
  \bibnamefont {Reinecke}}, \bibinfo {author} {\bibfnamefont {S.~N.}\
  \bibnamefont {Walck}}, \bibinfo {author} {\bibfnamefont {J.~P.}\ \bibnamefont
  {Reithmaier}}, \bibinfo {author} {\bibfnamefont {F.}~\bibnamefont {Klopf}}, \
  and\ \bibinfo {author} {\bibfnamefont {F.}~\bibnamefont {Sch\"afer}},\ }\href
  {\doibase 10.1103/PhysRevB.65.195315} {\bibfield  {journal} {\bibinfo
  {journal} {Phys. Rev. B}\ }\textbf {\bibinfo {volume} {65}},\ \bibinfo
  {pages} {195315} (\bibinfo {year} {2002})}\BibitemShut {NoStop}%
\bibitem [{\citenamefont {Tholen}\ \emph {et~al.}(2016)\citenamefont {Tholen},
  \citenamefont {Wildmann}, \citenamefont {Rastelli}, \citenamefont {Trotta},
  \citenamefont {Pryor}, \citenamefont {Zallo}, \citenamefont {Schmidt},
  \citenamefont {Koenraad},\ and\ \citenamefont {Silov}}]{PhysRevB.94.245301}%
  \BibitemOpen
  \bibfield  {author} {\bibinfo {author} {\bibfnamefont {H.~M. G.~A.}\
  \bibnamefont {Tholen}}, \bibinfo {author} {\bibfnamefont {J.~S.}\
  \bibnamefont {Wildmann}}, \bibinfo {author} {\bibfnamefont {A.}~\bibnamefont
  {Rastelli}}, \bibinfo {author} {\bibfnamefont {R.}~\bibnamefont {Trotta}},
  \bibinfo {author} {\bibfnamefont {C.~E.}\ \bibnamefont {Pryor}}, \bibinfo
  {author} {\bibfnamefont {E.}~\bibnamefont {Zallo}}, \bibinfo {author}
  {\bibfnamefont {O.~G.}\ \bibnamefont {Schmidt}}, \bibinfo {author}
  {\bibfnamefont {P.~M.}\ \bibnamefont {Koenraad}}, \ and\ \bibinfo {author}
  {\bibfnamefont {A.~Y.}\ \bibnamefont {Silov}},\ }\href {\doibase
  10.1103/PhysRevB.94.245301} {\bibfield  {journal} {\bibinfo  {journal} {Phys.
  Rev. B}\ }\textbf {\bibinfo {volume} {94}},\ \bibinfo {pages} {245301}
  (\bibinfo {year} {2016})}\BibitemShut {NoStop}%
\bibitem [{\citenamefont {Sheng}(2010)}]{doi:10.1063/1.3367707}%
  \BibitemOpen
  \bibfield  {author} {\bibinfo {author} {\bibfnamefont {W.}~\bibnamefont
  {Sheng}},\ }\href {\doibase 10.1063/1.3367707} {\bibfield  {journal}
  {\bibinfo  {journal} {Applied Physics Letters}\ }\textbf {\bibinfo {volume}
  {96}},\ \bibinfo {pages} {133102} (\bibinfo {year} {2010})},\ \Eprint
  {http://arxiv.org/abs/https://doi.org/10.1063/1.3367707}
  {https://doi.org/10.1063/1.3367707} \BibitemShut {NoStop}%
\bibitem [{\citenamefont {Witek}\ \emph {et~al.}(2011)\citenamefont {Witek},
  \citenamefont {Heeres}, \citenamefont {Perinetti}, \citenamefont {Bakkers},
  \citenamefont {Kouwenhoven},\ and\ \citenamefont
  {Zwiller}}]{PhysRevB.84.195305}%
  \BibitemOpen
  \bibfield  {author} {\bibinfo {author} {\bibfnamefont {B.~J.}\ \bibnamefont
  {Witek}}, \bibinfo {author} {\bibfnamefont {R.~W.}\ \bibnamefont {Heeres}},
  \bibinfo {author} {\bibfnamefont {U.}~\bibnamefont {Perinetti}}, \bibinfo
  {author} {\bibfnamefont {E.~P. A.~M.}\ \bibnamefont {Bakkers}}, \bibinfo
  {author} {\bibfnamefont {L.~P.}\ \bibnamefont {Kouwenhoven}}, \ and\ \bibinfo
  {author} {\bibfnamefont {V.}~\bibnamefont {Zwiller}},\ }\href {\doibase
  10.1103/PhysRevB.84.195305} {\bibfield  {journal} {\bibinfo  {journal} {Phys.
  Rev. B}\ }\textbf {\bibinfo {volume} {84}},\ \bibinfo {pages} {195305}
  (\bibinfo {year} {2011})}\BibitemShut {NoStop}%
\bibitem [{\citenamefont {Huo}\ \emph {et~al.}(2014)\citenamefont {Huo},
  \citenamefont {Witek}, \citenamefont {Kumar}, \citenamefont {Cardenas},
  \citenamefont {Zhang}, \citenamefont {Akopian}, \citenamefont {Singh},
  \citenamefont {Zallo}, \citenamefont {Grifone}, \citenamefont {Kriegner},
  \citenamefont {Trotta}, \citenamefont {Ding}, \citenamefont {Stangl},
  \citenamefont {Zwiller}, \citenamefont {Bester}, \citenamefont {Rastelli},\
  and\ \citenamefont {Schmidt}}]{Huo:NatPhys}%
  \BibitemOpen
  \bibfield  {author} {\bibinfo {author} {\bibfnamefont {Y.~H.}\ \bibnamefont
  {Huo}}, \bibinfo {author} {\bibfnamefont {B.~J.}\ \bibnamefont {Witek}},
  \bibinfo {author} {\bibfnamefont {S.}~\bibnamefont {Kumar}}, \bibinfo
  {author} {\bibfnamefont {J.~R.}\ \bibnamefont {Cardenas}}, \bibinfo {author}
  {\bibfnamefont {J.~X.}\ \bibnamefont {Zhang}}, \bibinfo {author}
  {\bibfnamefont {N.}~\bibnamefont {Akopian}}, \bibinfo {author} {\bibfnamefont
  {R.}~\bibnamefont {Singh}}, \bibinfo {author} {\bibfnamefont
  {E.}~\bibnamefont {Zallo}}, \bibinfo {author} {\bibfnamefont
  {R.}~\bibnamefont {Grifone}}, \bibinfo {author} {\bibfnamefont
  {D.}~\bibnamefont {Kriegner}}, \bibinfo {author} {\bibfnamefont
  {R.}~\bibnamefont {Trotta}}, \bibinfo {author} {\bibfnamefont
  {F.}~\bibnamefont {Ding}}, \bibinfo {author} {\bibfnamefont {J.}~\bibnamefont
  {Stangl}}, \bibinfo {author} {\bibfnamefont {V.}~\bibnamefont {Zwiller}},
  \bibinfo {author} {\bibfnamefont {G.}~\bibnamefont {Bester}}, \bibinfo
  {author} {\bibfnamefont {A.}~\bibnamefont {Rastelli}}, \ and\ \bibinfo
  {author} {\bibfnamefont {O.~G.}\ \bibnamefont {Schmidt}},\ }\href@noop {}
  {\bibfield  {journal} {\bibinfo  {journal} {Nature Physics}\ }\textbf
  {\bibinfo {volume} {10}},\ \bibinfo {pages} {46–51} (\bibinfo {year}
  {2014})}\BibitemShut {NoStop}%
\bibitem [{\citenamefont {Ulhaq}\ \emph {et~al.}(2016)\citenamefont {Ulhaq},
  \citenamefont {Duan}, \citenamefont {Zallo}, \citenamefont {Ding},
  \citenamefont {Schmidt}, \citenamefont {Tartakovskii}, \citenamefont
  {Skolnick},\ and\ \citenamefont {Chekhovich}}]{PhysRevB.93.165306}%
  \BibitemOpen
  \bibfield  {author} {\bibinfo {author} {\bibfnamefont {A.}~\bibnamefont
  {Ulhaq}}, \bibinfo {author} {\bibfnamefont {Q.}~\bibnamefont {Duan}},
  \bibinfo {author} {\bibfnamefont {E.}~\bibnamefont {Zallo}}, \bibinfo
  {author} {\bibfnamefont {F.}~\bibnamefont {Ding}}, \bibinfo {author}
  {\bibfnamefont {O.~G.}\ \bibnamefont {Schmidt}}, \bibinfo {author}
  {\bibfnamefont {A.~I.}\ \bibnamefont {Tartakovskii}}, \bibinfo {author}
  {\bibfnamefont {M.~S.}\ \bibnamefont {Skolnick}}, \ and\ \bibinfo {author}
  {\bibfnamefont {E.~A.}\ \bibnamefont {Chekhovich}},\ }\href {\doibase
  10.1103/PhysRevB.93.165306} {\bibfield  {journal} {\bibinfo  {journal} {Phys.
  Rev. B}\ }\textbf {\bibinfo {volume} {93}},\ \bibinfo {pages} {165306}
  (\bibinfo {year} {2016})}\BibitemShut {NoStop}%
\bibitem [{\citenamefont {L\"obl}\ \emph {et~al.}(2019)\citenamefont {L\"obl},
  \citenamefont {Zhai}, \citenamefont {Jahn}, \citenamefont {Ritzmann},
  \citenamefont {Huo}, \citenamefont {Wieck}, \citenamefont {Schmidt},
  \citenamefont {Ludwig}, \citenamefont {Rastelli},\ and\ \citenamefont
  {Warburton}}]{lobl2019}%
  \BibitemOpen
  \bibfield  {author} {\bibinfo {author} {\bibfnamefont {M.~C.}\ \bibnamefont
  {L\"obl}}, \bibinfo {author} {\bibfnamefont {L.}~\bibnamefont {Zhai}},
  \bibinfo {author} {\bibfnamefont {J.-P.}\ \bibnamefont {Jahn}}, \bibinfo
  {author} {\bibfnamefont {J.}~\bibnamefont {Ritzmann}}, \bibinfo {author}
  {\bibfnamefont {Y.}~\bibnamefont {Huo}}, \bibinfo {author} {\bibfnamefont
  {A.~D.}\ \bibnamefont {Wieck}}, \bibinfo {author} {\bibfnamefont {O.~G.}\
  \bibnamefont {Schmidt}}, \bibinfo {author} {\bibfnamefont {A.}~\bibnamefont
  {Ludwig}}, \bibinfo {author} {\bibfnamefont {A.}~\bibnamefont {Rastelli}}, \
  and\ \bibinfo {author} {\bibfnamefont {R.~J.}\ \bibnamefont {Warburton}},\
  }\href@noop {} {\bibfield  {journal} {\bibinfo  {journal} {Preprint at:
  arXiv:1902.10145}\ } (\bibinfo {year} {2019})}\BibitemShut {NoStop}%
\bibitem [{\citenamefont {Takagahara}(1993)}]{Takagahara1993}%
  \BibitemOpen
  \bibfield  {author} {\bibinfo {author} {\bibfnamefont {T.}~\bibnamefont
  {Takagahara}},\ }\href {\doibase 10.1103/PhysRevB.47.4569} {\bibfield
  {journal} {\bibinfo  {journal} {Physical Review B}\ }\textbf {\bibinfo
  {volume} {47}},\ \bibinfo {pages} {4569} (\bibinfo {year}
  {1993})}\BibitemShut {NoStop}%
\bibitem [{\citenamefont {Shumway}\ \emph {et~al.}(2001)\citenamefont
  {Shumway}, \citenamefont {Franceschetti},\ and\ \citenamefont
  {Zunger}}]{SFZ01}%
  \BibitemOpen
  \bibfield  {author} {\bibinfo {author} {\bibfnamefont {J.}~\bibnamefont
  {Shumway}}, \bibinfo {author} {\bibfnamefont {A.}~\bibnamefont
  {Franceschetti}}, \ and\ \bibinfo {author} {\bibfnamefont {A.}~\bibnamefont
  {Zunger}},\ }\href@noop {} {\bibfield  {journal} {\bibinfo  {journal} {Phys.
  Rev. B}\ }\textbf {\bibinfo {volume} {63}},\ \bibinfo {pages} {155316}
  (\bibinfo {year} {2001})}\BibitemShut {NoStop}%
\bibitem [{\citenamefont {Schliwa}\ \emph {et~al.}(2009)\citenamefont
  {Schliwa}, \citenamefont {Winkelnkemper},\ and\ \citenamefont
  {Bimberg}}]{Schliwa:09}%
  \BibitemOpen
  \bibfield  {author} {\bibinfo {author} {\bibfnamefont {A.}~\bibnamefont
  {Schliwa}}, \bibinfo {author} {\bibfnamefont {M.}~\bibnamefont
  {Winkelnkemper}}, \ and\ \bibinfo {author} {\bibfnamefont {D.}~\bibnamefont
  {Bimberg}},\ }\href {\doibase 10.1103/PhysRevB.79.075443} {\bibfield
  {journal} {\bibinfo  {journal} {Phys. Rev. B}\ }\textbf {\bibinfo {volume}
  {79}},\ \bibinfo {pages} {075443} (\bibinfo {year} {2009})}\BibitemShut
  {NoStop}%
\bibitem [{\citenamefont {Klenovsk\'y}\ \emph {et~al.}(2017)\citenamefont
  {Klenovsk\'y}, \citenamefont {Steindl},\ and\ \citenamefont
  {Geffroy}}]{Klenovsky2017}%
  \BibitemOpen
  \bibfield  {author} {\bibinfo {author} {\bibfnamefont {P.}~\bibnamefont
  {Klenovsk\'y}}, \bibinfo {author} {\bibfnamefont {P.}~\bibnamefont
  {Steindl}}, \ and\ \bibinfo {author} {\bibfnamefont {D.}~\bibnamefont
  {Geffroy}},\ }\href {\doibase 10.1038/srep45568} {\bibfield  {journal}
  {\bibinfo  {journal} {Scientific Reports}\ }\textbf {\bibinfo {volume} {7}},\
  \bibinfo {pages} {45568} (\bibinfo {year} {2017})}\BibitemShut {NoStop}%
\bibitem [{\citenamefont {Kunz}(2013)}]{Kunz2013}%
  \BibitemOpen
  \bibfield  {author} {\bibinfo {author} {\bibfnamefont {S.}~\bibnamefont
  {Kunz}},\ }\href@noop {} {\bibfield  {journal} {\bibinfo  {journal}
  {Magneto-optical properties of individual GaAs/AlGaAs Quantum Dots grown by
  Droplet Epitaxy, Institut national des sciences appliquées de Toulouse}\ }
  (\bibinfo {year} {2013})}\BibitemShut {NoStop}%
\bibitem [{\citenamefont {Cade}\ \emph {et~al.}(2006)\citenamefont {Cade},
  \citenamefont {Gotoh}, \citenamefont {Kamada}, \citenamefont {Nakano},\ and\
  \citenamefont {Okamoto}}]{PhysRevB.73.115322}%
  \BibitemOpen
  \bibfield  {author} {\bibinfo {author} {\bibfnamefont {N.~I.}\ \bibnamefont
  {Cade}}, \bibinfo {author} {\bibfnamefont {H.}~\bibnamefont {Gotoh}},
  \bibinfo {author} {\bibfnamefont {H.}~\bibnamefont {Kamada}}, \bibinfo
  {author} {\bibfnamefont {H.}~\bibnamefont {Nakano}}, \ and\ \bibinfo {author}
  {\bibfnamefont {H.}~\bibnamefont {Okamoto}},\ }\href {\doibase
  10.1103/PhysRevB.73.115322} {\bibfield  {journal} {\bibinfo  {journal} {Phys.
  Rev. B}\ }\textbf {\bibinfo {volume} {73}},\ \bibinfo {pages} {115322}
  (\bibinfo {year} {2006})}\BibitemShut {NoStop}%
\bibitem [{\citenamefont {Kavokin}(2003)}]{doi:10.1002/pssa.200306157}%
  \BibitemOpen
  \bibfield  {author} {\bibinfo {author} {\bibfnamefont {K.~V.}\ \bibnamefont
  {Kavokin}},\ }\href {\doibase 10.1002/pssa.200306157} {\bibfield  {journal}
  {\bibinfo  {journal} {physica status solidi (a)}\ }\textbf {\bibinfo {volume}
  {195}},\ \bibinfo {pages} {592} (\bibinfo {year} {2003})}\BibitemShut
  {NoStop}%
\bibitem [{\citenamefont {Igarashi}\ \emph {et~al.}(2010)\citenamefont
  {Igarashi}, \citenamefont {Shirane}, \citenamefont {Ota}, \citenamefont
  {Nomura}, \citenamefont {Kumagai}, \citenamefont {Ohkouchi}, \citenamefont
  {Kirihara}, \citenamefont {Ishida}, \citenamefont {Iwamoto}, \citenamefont
  {Yorozu},\ and\ \citenamefont {Arakawa}}]{PhysRevB.81.245304}%
  \BibitemOpen
  \bibfield  {author} {\bibinfo {author} {\bibfnamefont {Y.}~\bibnamefont
  {Igarashi}}, \bibinfo {author} {\bibfnamefont {M.}~\bibnamefont {Shirane}},
  \bibinfo {author} {\bibfnamefont {Y.}~\bibnamefont {Ota}}, \bibinfo {author}
  {\bibfnamefont {M.}~\bibnamefont {Nomura}}, \bibinfo {author} {\bibfnamefont
  {N.}~\bibnamefont {Kumagai}}, \bibinfo {author} {\bibfnamefont
  {S.}~\bibnamefont {Ohkouchi}}, \bibinfo {author} {\bibfnamefont
  {A.}~\bibnamefont {Kirihara}}, \bibinfo {author} {\bibfnamefont
  {S.}~\bibnamefont {Ishida}}, \bibinfo {author} {\bibfnamefont
  {S.}~\bibnamefont {Iwamoto}}, \bibinfo {author} {\bibfnamefont
  {S.}~\bibnamefont {Yorozu}}, \ and\ \bibinfo {author} {\bibfnamefont
  {Y.}~\bibnamefont {Arakawa}},\ }\href {\doibase 10.1103/PhysRevB.81.245304}
  {\bibfield  {journal} {\bibinfo  {journal} {Phys. Rev. B}\ }\textbf {\bibinfo
  {volume} {81}},\ \bibinfo {pages} {245304} (\bibinfo {year}
  {2010})}\BibitemShut {NoStop}%
\bibitem [{\citenamefont {Warming}\ \emph {et~al.}(2009)\citenamefont
  {Warming}, \citenamefont {Siebert}, \citenamefont {Schliwa}, \citenamefont
  {Stock}, \citenamefont {Zimmermann},\ and\ \citenamefont
  {Bimberg}}]{PhysRevB.79.125316}%
  \BibitemOpen
  \bibfield  {author} {\bibinfo {author} {\bibfnamefont {T.}~\bibnamefont
  {Warming}}, \bibinfo {author} {\bibfnamefont {E.}~\bibnamefont {Siebert}},
  \bibinfo {author} {\bibfnamefont {A.}~\bibnamefont {Schliwa}}, \bibinfo
  {author} {\bibfnamefont {E.}~\bibnamefont {Stock}}, \bibinfo {author}
  {\bibfnamefont {R.}~\bibnamefont {Zimmermann}}, \ and\ \bibinfo {author}
  {\bibfnamefont {D.}~\bibnamefont {Bimberg}},\ }\href {\doibase
  10.1103/PhysRevB.79.125316} {\bibfield  {journal} {\bibinfo  {journal} {Phys.
  Rev. B}\ }\textbf {\bibinfo {volume} {79}},\ \bibinfo {pages} {125316}
  (\bibinfo {year} {2009})}\BibitemShut {NoStop}%
\bibitem [{\citenamefont {Bayer}\ \emph {et~al.}(1998)\citenamefont {Bayer},
  \citenamefont {Walck}, \citenamefont {Reinecke},\ and\ \citenamefont
  {Forchel}}]{PhysRevB.57.6584}%
  \BibitemOpen
  \bibfield  {author} {\bibinfo {author} {\bibfnamefont {M.}~\bibnamefont
  {Bayer}}, \bibinfo {author} {\bibfnamefont {S.~N.}\ \bibnamefont {Walck}},
  \bibinfo {author} {\bibfnamefont {T.~L.}\ \bibnamefont {Reinecke}}, \ and\
  \bibinfo {author} {\bibfnamefont {A.}~\bibnamefont {Forchel}},\ }\href
  {\doibase 10.1103/PhysRevB.57.6584} {\bibfield  {journal} {\bibinfo
  {journal} {Phys. Rev. B}\ }\textbf {\bibinfo {volume} {57}},\ \bibinfo
  {pages} {6584} (\bibinfo {year} {1998})}\BibitemShut {NoStop}%
\bibitem [{\citenamefont {Schulhauser}\ \emph {et~al.}(2002)\citenamefont
  {Schulhauser}, \citenamefont {Haft}, \citenamefont {Warburton}, \citenamefont
  {Karrai}, \citenamefont {Govorov}, \citenamefont {Kalameitsev}, \citenamefont
  {Chaplik}, \citenamefont {Schoenfeld}, \citenamefont {Garcia},\ and\
  \citenamefont {Petroff}}]{PhysRevB.66.193303}%
  \BibitemOpen
  \bibfield  {author} {\bibinfo {author} {\bibfnamefont {C.}~\bibnamefont
  {Schulhauser}}, \bibinfo {author} {\bibfnamefont {D.}~\bibnamefont {Haft}},
  \bibinfo {author} {\bibfnamefont {R.~J.}\ \bibnamefont {Warburton}}, \bibinfo
  {author} {\bibfnamefont {K.}~\bibnamefont {Karrai}}, \bibinfo {author}
  {\bibfnamefont {A.~O.}\ \bibnamefont {Govorov}}, \bibinfo {author}
  {\bibfnamefont {A.~V.}\ \bibnamefont {Kalameitsev}}, \bibinfo {author}
  {\bibfnamefont {A.}~\bibnamefont {Chaplik}}, \bibinfo {author} {\bibfnamefont
  {W.}~\bibnamefont {Schoenfeld}}, \bibinfo {author} {\bibfnamefont {J.~M.}\
  \bibnamefont {Garcia}}, \ and\ \bibinfo {author} {\bibfnamefont {P.~M.}\
  \bibnamefont {Petroff}},\ }\href {\doibase 10.1103/PhysRevB.66.193303}
  {\bibfield  {journal} {\bibinfo  {journal} {Phys. Rev. B}\ }\textbf {\bibinfo
  {volume} {66}},\ \bibinfo {pages} {193303} (\bibinfo {year}
  {2002})}\BibitemShut {NoStop}%
\bibitem [{\citenamefont {Tsai}\ \emph {et~al.}(2008)\citenamefont {Tsai},
  \citenamefont {Lin}, \citenamefont {Lin}, \citenamefont {Lin}, \citenamefont
  {Wang}, \citenamefont {Lo}, \citenamefont {Cheng}, \citenamefont {Lee},\ and\
  \citenamefont {Chang}}]{PhysRevLett.101.267402}%
  \BibitemOpen
  \bibfield  {author} {\bibinfo {author} {\bibfnamefont {M.-F.}\ \bibnamefont
  {Tsai}}, \bibinfo {author} {\bibfnamefont {H.}~\bibnamefont {Lin}}, \bibinfo
  {author} {\bibfnamefont {C.-H.}\ \bibnamefont {Lin}}, \bibinfo {author}
  {\bibfnamefont {S.-D.}\ \bibnamefont {Lin}}, \bibinfo {author} {\bibfnamefont
  {S.-Y.}\ \bibnamefont {Wang}}, \bibinfo {author} {\bibfnamefont {M.-C.}\
  \bibnamefont {Lo}}, \bibinfo {author} {\bibfnamefont {S.-J.}\ \bibnamefont
  {Cheng}}, \bibinfo {author} {\bibfnamefont {M.-C.}\ \bibnamefont {Lee}}, \
  and\ \bibinfo {author} {\bibfnamefont {W.-H.}\ \bibnamefont {Chang}},\ }\href
  {\doibase 10.1103/PhysRevLett.101.267402} {\bibfield  {journal} {\bibinfo
  {journal} {Phys. Rev. Lett.}\ }\textbf {\bibinfo {volume} {101}},\ \bibinfo
  {pages} {267402} (\bibinfo {year} {2008})}\BibitemShut {NoStop}%
\bibitem [{\citenamefont {Fu}\ \emph {et~al.}(2010)\citenamefont {Fu},
  \citenamefont {Lin}, \citenamefont {Tsai}, \citenamefont {Lin}, \citenamefont
  {Lin}, \citenamefont {Chou}, \citenamefont {Cheng},\ and\ \citenamefont
  {Chang}}]{PhysRevB.81.113307}%
  \BibitemOpen
  \bibfield  {author} {\bibinfo {author} {\bibfnamefont {Y.~J.}\ \bibnamefont
  {Fu}}, \bibinfo {author} {\bibfnamefont {S.~D.}\ \bibnamefont {Lin}},
  \bibinfo {author} {\bibfnamefont {M.~F.}\ \bibnamefont {Tsai}}, \bibinfo
  {author} {\bibfnamefont {H.}~\bibnamefont {Lin}}, \bibinfo {author}
  {\bibfnamefont {C.~H.}\ \bibnamefont {Lin}}, \bibinfo {author} {\bibfnamefont
  {H.~Y.}\ \bibnamefont {Chou}}, \bibinfo {author} {\bibfnamefont {S.~J.}\
  \bibnamefont {Cheng}}, \ and\ \bibinfo {author} {\bibfnamefont {W.~H.}\
  \bibnamefont {Chang}},\ }\href {\doibase 10.1103/PhysRevB.81.113307}
  {\bibfield  {journal} {\bibinfo  {journal} {Phys. Rev. B}\ }\textbf {\bibinfo
  {volume} {81}},\ \bibinfo {pages} {113307} (\bibinfo {year}
  {2010})}\BibitemShut {NoStop}%
\bibitem [{\citenamefont {van Bree}\ \emph {et~al.}(2012)\citenamefont {van
  Bree}, \citenamefont {Silov}, \citenamefont {Koenraad}, \citenamefont
  {Flatt\'e},\ and\ \citenamefont {Pryor}}]{PhysRevB.85.165323}%
  \BibitemOpen
  \bibfield  {author} {\bibinfo {author} {\bibfnamefont {J.}~\bibnamefont {van
  Bree}}, \bibinfo {author} {\bibfnamefont {A.~Y.}\ \bibnamefont {Silov}},
  \bibinfo {author} {\bibfnamefont {P.~M.}\ \bibnamefont {Koenraad}}, \bibinfo
  {author} {\bibfnamefont {M.~E.}\ \bibnamefont {Flatt\'e}}, \ and\ \bibinfo
  {author} {\bibfnamefont {C.~E.}\ \bibnamefont {Pryor}},\ }\href {\doibase
  10.1103/PhysRevB.85.165323} {\bibfield  {journal} {\bibinfo  {journal} {Phys.
  Rev. B}\ }\textbf {\bibinfo {volume} {85}},\ \bibinfo {pages} {165323}
  (\bibinfo {year} {2012})}\BibitemShut {NoStop}%
\bibitem [{\citenamefont {van Bree}\ \emph {et~al.}(2016)\citenamefont {van
  Bree}, \citenamefont {Silov}, \citenamefont {van Maasakkers}, \citenamefont
  {Pryor}, \citenamefont {Flatt\'e},\ and\ \citenamefont
  {Koenraad}}]{PhysRevB.93.035311}%
  \BibitemOpen
  \bibfield  {author} {\bibinfo {author} {\bibfnamefont {J.}~\bibnamefont {van
  Bree}}, \bibinfo {author} {\bibfnamefont {A.~Y.}\ \bibnamefont {Silov}},
  \bibinfo {author} {\bibfnamefont {M.~L.}\ \bibnamefont {van Maasakkers}},
  \bibinfo {author} {\bibfnamefont {C.~E.}\ \bibnamefont {Pryor}}, \bibinfo
  {author} {\bibfnamefont {M.~E.}\ \bibnamefont {Flatt\'e}}, \ and\ \bibinfo
  {author} {\bibfnamefont {P.~M.}\ \bibnamefont {Koenraad}},\ }\href@noop {}
  {\bibfield  {journal} {\bibinfo  {journal} {Phys. Rev. B}\ }\textbf {\bibinfo
  {volume} {93}},\ \bibinfo {pages} {035311} (\bibinfo {year}
  {2016})}\BibitemShut {NoStop}%
\bibitem [{\citenamefont {Schwan}\ \emph {et~al.}(2011)\citenamefont {Schwan},
  \citenamefont {Meiners}, \citenamefont {Greilich}, \citenamefont {Yakovlev},
  \citenamefont {Bayer}, \citenamefont {Maia}, \citenamefont {Quivy},\ and\
  \citenamefont {Henriques}}]{doi:10.1063/1.3665634}%
  \BibitemOpen
  \bibfield  {author} {\bibinfo {author} {\bibfnamefont {A.}~\bibnamefont
  {Schwan}}, \bibinfo {author} {\bibfnamefont {B.-M.}\ \bibnamefont {Meiners}},
  \bibinfo {author} {\bibfnamefont {A.}~\bibnamefont {Greilich}}, \bibinfo
  {author} {\bibfnamefont {D.~R.}\ \bibnamefont {Yakovlev}}, \bibinfo {author}
  {\bibfnamefont {M.}~\bibnamefont {Bayer}}, \bibinfo {author} {\bibfnamefont
  {A.~D.~B.}\ \bibnamefont {Maia}}, \bibinfo {author} {\bibfnamefont {A.~A.}\
  \bibnamefont {Quivy}}, \ and\ \bibinfo {author} {\bibfnamefont {A.~B.}\
  \bibnamefont {Henriques}},\ }\href {\doibase 10.1063/1.3665634} {\bibfield
  {journal} {\bibinfo  {journal} {Applied Physics Letters}\ }\textbf {\bibinfo
  {volume} {99}},\ \bibinfo {pages} {221914} (\bibinfo {year}
  {2011})}\BibitemShut {NoStop}%
\bibitem [{\citenamefont {Pryor}\ and\ \citenamefont
  {Flatt\'e}(2006)}]{PhysRevLett.96.026804}%
  \BibitemOpen
  \bibfield  {author} {\bibinfo {author} {\bibfnamefont {C.~E.}\ \bibnamefont
  {Pryor}}\ and\ \bibinfo {author} {\bibfnamefont {M.~E.}\ \bibnamefont
  {Flatt\'e}},\ }\href {\doibase 10.1103/PhysRevLett.96.026804} {\bibfield
  {journal} {\bibinfo  {journal} {Phys. Rev. Lett.}\ }\textbf {\bibinfo
  {volume} {96}},\ \bibinfo {pages} {026804} (\bibinfo {year}
  {2006})}\BibitemShut {NoStop}%
\bibitem [{\citenamefont {Watzinger}\ \emph {et~al.}(2016)\citenamefont
  {Watzinger}, \citenamefont {Kloeffel}, \citenamefont {Vukusic}, \citenamefont
  {Rossell}, \citenamefont {Sessi}, \citenamefont {Kukucka}, \citenamefont
  {Kirchschlager}, \citenamefont {Lausecker}, \citenamefont {Truhlar},
  \citenamefont {Glaser}, \citenamefont {Rastelli}, \citenamefont {Fuhrer},
  \citenamefont {Loss},\ and\ \citenamefont
  {Katsaros}}]{10.1021/acs.nanolett.6b02715}%
  \BibitemOpen
  \bibfield  {author} {\bibinfo {author} {\bibfnamefont {H.}~\bibnamefont
  {Watzinger}}, \bibinfo {author} {\bibfnamefont {C.}~\bibnamefont {Kloeffel}},
  \bibinfo {author} {\bibfnamefont {L.}~\bibnamefont {Vukusic}}, \bibinfo
  {author} {\bibfnamefont {M.~D.}\ \bibnamefont {Rossell}}, \bibinfo {author}
  {\bibfnamefont {V.}~\bibnamefont {Sessi}}, \bibinfo {author} {\bibfnamefont
  {J.}~\bibnamefont {Kukucka}}, \bibinfo {author} {\bibfnamefont
  {R.}~\bibnamefont {Kirchschlager}}, \bibinfo {author} {\bibfnamefont
  {E.}~\bibnamefont {Lausecker}}, \bibinfo {author} {\bibfnamefont
  {A.}~\bibnamefont {Truhlar}}, \bibinfo {author} {\bibfnamefont
  {M.}~\bibnamefont {Glaser}}, \bibinfo {author} {\bibfnamefont
  {A.}~\bibnamefont {Rastelli}}, \bibinfo {author} {\bibfnamefont
  {A.}~\bibnamefont {Fuhrer}}, \bibinfo {author} {\bibfnamefont
  {D.}~\bibnamefont {Loss}}, \ and\ \bibinfo {author} {\bibfnamefont
  {G.}~\bibnamefont {Katsaros}},\ }\href {\doibase
  10.1021/acs.nanolett.6b02715} {\bibfield  {journal} {\bibinfo  {journal}
  {Nano letters}\ }\textbf {\bibinfo {volume} {16}},\ \bibinfo {pages} {6879}
  (\bibinfo {year} {2016})}\BibitemShut {NoStop}%
\bibitem [{\citenamefont {Akimov}\ \emph {et~al.}(2005)\citenamefont {Akimov},
  \citenamefont {Kavokin}, \citenamefont {Hundt},\ and\ \citenamefont
  {Henneberger}}]{PhysRevB.71.075326}%
  \BibitemOpen
  \bibfield  {author} {\bibinfo {author} {\bibfnamefont {I.~A.}\ \bibnamefont
  {Akimov}}, \bibinfo {author} {\bibfnamefont {K.~V.}\ \bibnamefont {Kavokin}},
  \bibinfo {author} {\bibfnamefont {A.}~\bibnamefont {Hundt}}, \ and\ \bibinfo
  {author} {\bibfnamefont {F.}~\bibnamefont {Henneberger}},\ }\href {\doibase
  10.1103/PhysRevB.71.075326} {\bibfield  {journal} {\bibinfo  {journal} {Phys.
  Rev. B}\ }\textbf {\bibinfo {volume} {71}},\ \bibinfo {pages} {075326}
  (\bibinfo {year} {2005})}\BibitemShut {NoStop}%
\bibitem [{\citenamefont {Durnev}\ \emph {et~al.}(2012)\citenamefont {Durnev},
  \citenamefont {Glazov},\ and\ \citenamefont {Ivchenko}}]{DURNEV2012797}%
  \BibitemOpen
  \bibfield  {author} {\bibinfo {author} {\bibfnamefont {M.}~\bibnamefont
  {Durnev}}, \bibinfo {author} {\bibfnamefont {M.}~\bibnamefont {Glazov}}, \
  and\ \bibinfo {author} {\bibfnamefont {E.}~\bibnamefont {Ivchenko}},\ }\href
  {\doibase https://doi.org/10.1016/j.physe.2011.12.003} {\bibfield  {journal}
  {\bibinfo  {journal} {Physica E: Low-dimensional Systems and Nanostructures}\
  }\textbf {\bibinfo {volume} {44}},\ \bibinfo {pages} {797 } (\bibinfo {year}
  {2012})}\BibitemShut {NoStop}%
\bibitem [{\citenamefont {Birner}\ \emph {et~al.}(2007)\citenamefont {Birner},
  \citenamefont {Zibold}, \citenamefont {Andlauer}, \citenamefont {Kubis},
  \citenamefont {Sabathil}, \citenamefont {Trellakis},\ and\ \citenamefont
  {Vogl}}]{Birner:07}%
  \BibitemOpen
  \bibfield  {author} {\bibinfo {author} {\bibfnamefont {S.}~\bibnamefont
  {Birner}}, \bibinfo {author} {\bibfnamefont {T.}~\bibnamefont {Zibold}},
  \bibinfo {author} {\bibfnamefont {T.}~\bibnamefont {Andlauer}}, \bibinfo
  {author} {\bibfnamefont {T.}~\bibnamefont {Kubis}}, \bibinfo {author}
  {\bibfnamefont {M.}~\bibnamefont {Sabathil}}, \bibinfo {author}
  {\bibfnamefont {A.}~\bibnamefont {Trellakis}}, \ and\ \bibinfo {author}
  {\bibfnamefont {P.}~\bibnamefont {Vogl}},\ }\href@noop {} {\bibfield
  {journal} {\bibinfo  {journal} {IEEE Trans. El. Dev.}\ }\textbf {\bibinfo
  {volume} {54}},\ \bibinfo {pages} {2137} (\bibinfo {year}
  {2007})}\BibitemShut {NoStop}%
\bibitem [{Arx()}]{ArxivKlenovsky:19}%
  \BibitemOpen
  \href@noop {} {}\bibinfo {howpublished} {Petr Klenovsk\'y, Andrei Schliwa,
  Dieter Bimberg, Phys. Rev. B {\it in print}, arXiv:1903.09078}\BibitemShut
  {NoStop}%
\bibitem [{\citenamefont {Beya-Wakata}\ \emph {et~al.}(2011)\citenamefont
  {Beya-Wakata}, \citenamefont {Prodhomme},\ and\ \citenamefont
  {Bester}}]{Beya-Wakata2011}%
  \BibitemOpen
  \bibfield  {author} {\bibinfo {author} {\bibfnamefont {A.}~\bibnamefont
  {Beya-Wakata}}, \bibinfo {author} {\bibfnamefont {P.~Y.}\ \bibnamefont
  {Prodhomme}}, \ and\ \bibinfo {author} {\bibfnamefont {G.}~\bibnamefont
  {Bester}},\ }\href {\doibase 10.1103/PhysRevB.84.195207} {\bibfield
  {journal} {\bibinfo  {journal} {Physical Review B}\ }\textbf {\bibinfo
  {volume} {84}},\ \bibinfo {pages} {195207} (\bibinfo {year}
  {2011})}\BibitemShut {NoStop}%
\bibitem [{\citenamefont {Klenovsk\'y}\ \emph {et~al.}(2018)\citenamefont
  {Klenovsk\'y}, \citenamefont {Steindl}, \citenamefont {Aberl}, \citenamefont
  {Zallo}, \citenamefont {Trotta}, \citenamefont {Rastelli},\ and\
  \citenamefont {Fromherz}}]{Klenovsky2018}%
  \BibitemOpen
  \bibfield  {author} {\bibinfo {author} {\bibfnamefont {P.}~\bibnamefont
  {Klenovsk\'y}}, \bibinfo {author} {\bibfnamefont {P.}~\bibnamefont
  {Steindl}}, \bibinfo {author} {\bibfnamefont {J.}~\bibnamefont {Aberl}},
  \bibinfo {author} {\bibfnamefont {E.}~\bibnamefont {Zallo}}, \bibinfo
  {author} {\bibfnamefont {R.}~\bibnamefont {Trotta}}, \bibinfo {author}
  {\bibfnamefont {A.}~\bibnamefont {Rastelli}}, \ and\ \bibinfo {author}
  {\bibfnamefont {T.}~\bibnamefont {Fromherz}},\ }\href {\doibase
  10.1103/PhysRevB.97.245314} {\bibfield  {journal} {\bibinfo  {journal}
  {Physical Review B}\ }\textbf {\bibinfo {volume} {97}},\ \bibinfo {pages}
  {245314} (\bibinfo {year} {2018})}\BibitemShut {NoStop}%
\bibitem [{\citenamefont {Aberl}\ \emph {et~al.}(2017)\citenamefont {Aberl},
  \citenamefont {Klenovsk\'y}, \citenamefont {Wildmann}, \citenamefont
  {Mart\'in-S\'anchez}, \citenamefont {Fromherz}, \citenamefont {Zallo},
  \citenamefont {Huml\'i\v{c}ek}, \citenamefont {Rastelli},\ and\ \citenamefont
  {Trotta}}]{Aberl:17}%
  \BibitemOpen
  \bibfield  {author} {\bibinfo {author} {\bibfnamefont {J.}~\bibnamefont
  {Aberl}}, \bibinfo {author} {\bibfnamefont {P.}~\bibnamefont {Klenovsk\'y}},
  \bibinfo {author} {\bibfnamefont {J.~S.}\ \bibnamefont {Wildmann}}, \bibinfo
  {author} {\bibfnamefont {J.}~\bibnamefont {Mart\'in-S\'anchez}}, \bibinfo
  {author} {\bibfnamefont {T.}~\bibnamefont {Fromherz}}, \bibinfo {author}
  {\bibfnamefont {E.}~\bibnamefont {Zallo}}, \bibinfo {author} {\bibfnamefont
  {J.}~\bibnamefont {Huml\'i\v{c}ek}}, \bibinfo {author} {\bibfnamefont
  {A.}~\bibnamefont {Rastelli}}, \ and\ \bibinfo {author} {\bibfnamefont
  {R.}~\bibnamefont {Trotta}},\ }\href {\doibase 10.1103/PhysRevB.70.201308}
  {\bibfield  {journal} {\bibinfo  {journal} {Phys. Rev. B}\ }\textbf {\bibinfo
  {volume} {70}},\ \bibinfo {pages} {201308} (\bibinfo {year}
  {2017})}\BibitemShut {NoStop}%
\bibitem [{\citenamefont {Maialle}\ and\ \citenamefont
  {Degani}(2007)}]{Maialle:07}%
  \BibitemOpen
  \bibfield  {author} {\bibinfo {author} {\bibfnamefont {M.~Z.}\ \bibnamefont
  {Maialle}}\ and\ \bibinfo {author} {\bibfnamefont {M.~H.}\ \bibnamefont
  {Degani}},\ }\href {\doibase 10.1103/PhysRevB.76.115302} {\bibfield
  {journal} {\bibinfo  {journal} {Phys. Rev. B}\ }\textbf {\bibinfo {volume}
  {76}},\ \bibinfo {pages} {115302} (\bibinfo {year} {2007})}\BibitemShut
  {NoStop}%
\bibitem [{\citenamefont {Wang}\ \emph {et~al.}(2009)\citenamefont {Wang},
  \citenamefont {K\ifmmode~\check{r}\else \v{r}\fi{}\'apek}, \citenamefont
  {Ding}, \citenamefont {Horton}, \citenamefont {Schliwa}, \citenamefont
  {Bimberg}, \citenamefont {Rastelli},\ and\ \citenamefont
  {Schmidt}}]{PhysRevB.80.085309}%
  \BibitemOpen
  \bibfield  {author} {\bibinfo {author} {\bibfnamefont {L.}~\bibnamefont
  {Wang}}, \bibinfo {author} {\bibfnamefont {V.}~\bibnamefont
  {K\ifmmode~\check{r}\else \v{r}\fi{}\'apek}}, \bibinfo {author}
  {\bibfnamefont {F.}~\bibnamefont {Ding}}, \bibinfo {author} {\bibfnamefont
  {F.}~\bibnamefont {Horton}}, \bibinfo {author} {\bibfnamefont
  {A.}~\bibnamefont {Schliwa}}, \bibinfo {author} {\bibfnamefont
  {D.}~\bibnamefont {Bimberg}}, \bibinfo {author} {\bibfnamefont
  {A.}~\bibnamefont {Rastelli}}, \ and\ \bibinfo {author} {\bibfnamefont
  {O.~G.}\ \bibnamefont {Schmidt}},\ }\href {\doibase
  10.1103/PhysRevB.80.085309} {\bibfield  {journal} {\bibinfo  {journal} {Phys.
  Rev. B}\ }\textbf {\bibinfo {volume} {80}},\ \bibinfo {pages} {085309}
  (\bibinfo {year} {2009})}\BibitemShut {NoStop}%
\end{thebibliography}%

\end{document}